\documentclass[journal=jpchax,manuscript=article]{achemso}
\usepackage{achemso}
\usepackage{chemformula} % Formula subscripts using \ch{}
\usepackage[T1]{fontenc} % Use modern font encodings
\usepackage{amsmath}
\usepackage{amssymb}
\usepackage{color}
\usepackage{graphicx}
\usepackage{tabularx}
\usepackage{dcolumn}
\usepackage{bm}      % bold math
\usepackage{array}   % table making
\usepackage{amsfonts}
\usepackage{url}
\usepackage{xr-hyper}
\usepackage[bookmarks=false,colorlinks,citecolor=red]{hyperref}
\usepackage[utf8]{inputenc}
\usepackage[T1]{fontenc}
\usepackage{mathptmx}
\usepackage{multirow}
\usepackage{chemformula}
\usepackage{upgreek}
\usepackage{float}
\usepackage{subcaption}
\usepackage{cleveref}
\usepackage[font=small,labelfont=bf, justification=justified, format=plain]{caption}

\setkeys{acs}{articletitle = true}
\setkeys{acs}{maxauthors=10}
\setkeys{acs}{etalmode=truncate}

\newcommand{\CC}[1]{{\color{black}{#1}}}
\author{Cancan Huang}
\affiliation{Department of Chemistry, Brown University, Providence, Rhode Island, 02912, USA}

\author{Brenda M. Rubenstein}
\affiliation{Department of Chemistry, Brown University, Providence, Rhode Island, 02912, USA}
\email{brenda_rubenstein@brown.edu}

\title{Machine Learning Diffusion Monte Carlo Forces}

\keywords{Forces, Machine Learning, Quantum Monte Carlo, Diffusion Monte Carlo, Density Functional Theory}

\begin{document}

\begin{tocentry}
  \includegraphics{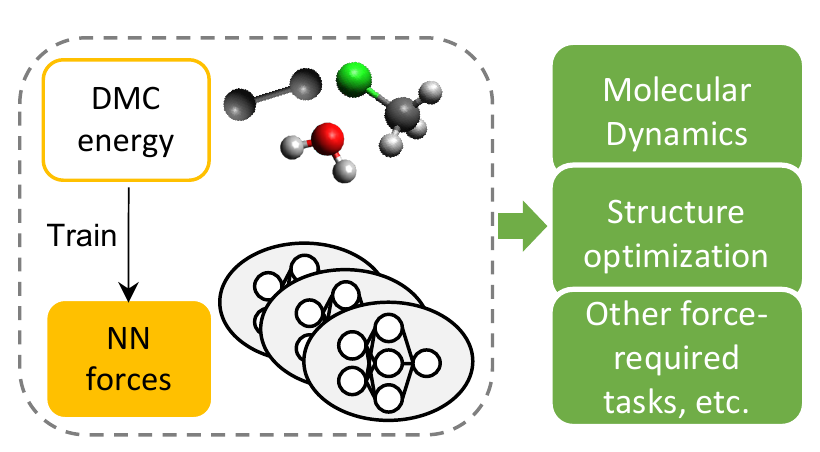}
\end{tocentry}

\begin{abstract}
  Diffusion Monte Carlo (DMC) is one of the most accurate techniques available for calculating the electronic properties of molecules and materials, yet it often remains a challenge to economically compute forces using this technique. As a result, \textit{ab initio} molecular dynamics simulations and geometry optimizations that employ Diffusion Monte Carlo forces are often out of reach. One potential approach for accelerating the computation of ``DMC forces'' is to machine learn these forces from DMC energy calculations. In this work, we employ Behler-Parrinello Neural Networks to learn DMC forces from DMC energy calculations for geometry optimization and molecular dynamics simulations of small molecules. We illustrate the unique challenges that stem from learning forces without explicit force data and from noisy energy data by making rigorous comparisons of potential energy surface, dynamics, and optimization predictions among \textit{ab initio} Density Functional Theory (DFT) simulations and machine learning models trained on DFT energies with forces, DFT energies without forces, and DMC energies without forces. We show for three small molecules - \ch{C2}, \ch{H2O}, and \ch{CH3Cl} - that machine learned DMC dynamics can reproduce average bond lengths and angles within a few percent of known experimental results at a 100th of the typical cost. Our work describes a much-needed means of performing dynamics simulations on high-accuracy, DMC PESs and for generating DMC-quality molecular geometries given current algorithmic constraints.
\end{abstract}

\section{Introduction}
\label{Introduction}
One of science's less-celebrated, yet revolutionary achievements has been the development of highly accurate electronic structure techniques, ranging from Density Functional Theory (DFT)\cite{kohn1996density} to Coupled Cluster Theory (CC)\cite{bishop1991overview} to Quantum Monte Carlo (QMC)\cite{foulkes2001quantum} methods, that enable researchers to accurately model chemical and biological processes with an unprecedented level of atomistic details. These methods allow scientists to `see' how molecules react,\cite{chenoweth2008reaxff} proteins fold,\cite{li2018finding} and electrons flow through materials,\cite{sprague2019maximizing} thus also granting them the powerful ability to engineer these molecules and materials atom-by-atom. While game-changing, these methods still have their limitations, chief among them being their computational cost. Because of their great expense, these techniques cannot be readily used to model systems comprised of thousands of atoms or more, meaning that they cannot be used to directly model complex or amorphous solids, multicomponent liquids, or very large macromolecular complexes. Indeed, even DFT, the cheapest fully \textit{ab initio} electronic structure technique, scales as $O(N^{3})$, where $N$ is the number of basis functions, and more accurate many-body techniques such as Moller-Plesset Perturbation Theory and CC can scale as steeply as $O(N^7)$ and $O(N^{10})$,\cite{bartlett2007coupled} respectively. In many cases, this limits the use of the highest accuracy techniques to comparatively small (less than 20-atom), gas-phase molecules. Exacerbating matters, for many electronic structure techniques, the calculation of forces is even more expensive than the calculation of energies. Since forces are needed to relax molecular and material geometries\cite{tiihonen2021toward, ly2022phonons} as well as to drive molecular dynamics (MD) simulations, this places the high-accuracy, \textit{ab initio} determination of molecular structures and dynamics out of practical reach for many techniques.

The computational overhead associated with calculating forces is especially problematic for many QMC techniques, such as the Diffusion Monte Carlo (DMC) method.\cite{kosztin1996introduction} Although QMC techniques are increasingly being used to provide computational ground truths regarding energies, wave functions, and electron densities in chemistry and condensed matter physics because of their exceptional accuracy\cite{melton2016spin, kylanpaa2017accuracy,tiihonen2021toward} and modest $O(N^{3})$-$O(N^{4})$ scaling,\cite{foulkes2001quantum} it has long been a challenge to calculate forces within most real-space QMC methods.\cite{hammond1994monte, assaraf2000computing,filippi2000correlated} This is because the direct application of the Hellman-Feynman Theorem,\cite{feynman1939forces} which is employed to compute forces in most theories, results in an infinite variance problem that stems from the sampling of electron positions that approach the nucleus.\cite{hammond1994monte,chiesa2005accurate} Moreover, most QMC techniques use a mixed estimator, $\langle \Phi_0 | \nabla \mathcal{H}| \Psi_T \rangle$, rather than a pure estimator, $\langle \Phi_0 | \nabla \mathcal{H}| \Phi_0 \rangle$, to compute forces and other observables, even though only the pure estimator produces an accurate estimate of forces.\cite{hammond1994monte, badinski2010methods} Zero-variance, zero-bias (ZVZB) estimators\cite{assaraf1999zero,assaraf2003zero,assaraf2000computing, badinski2008energy} and other advances\cite{badinski2007accurate, badinski2008total, badinski2008nodal, sorella2010algorithmic, attaccalite2008stable, ly2022phonons, moroni2014practical}, have been proposed that achieve significantly higher accuracy with greater statistical efficiency than the bare Hellman-Feynman estimator described above, but these techniques still suffer from biases and/or steep scalings that curtail their appeal and practicality.\cite{tiihonen2021toward} Most recently, several regularization techniques that eliminate the infinite variance problem have been proposed,\cite{pathak2020light, van2021energy,rios2019tail} yet arguments remain about their accuracy, scaling, and general utility in real systems. This state of affairs calls for the development of techniques that can accurately compute forces at the QMC level of theory, but at a substantially reduced cost. 

One appealing way to compute DMC forces at a reduced cost is via machine learning (ML) techniques. In recent years, machine learning techniques have been widely employed to accelerate the computation of energies,\cite{schmitz2019machine,bartok2010gaussian} electron densities,\cite{grisafi2018transferable,cuevas2021machine} and density matrices,\cite{cranmer2019inferring,wetherell2020insights} among other quantities, using such ML methods as Gaussian Process Regression (GPR),\cite{pmlr-v5-titsias09a} Kernel Ridge Regression (KRR),\cite{shawe2004kernel} Support Vector Regression (SVR),\cite{NIPS1996_d3890178} and Neural Networks (NNs).\cite{bishop1991overview} ML techniques are particularly attractive because, based upon a comparatively small set of training data, they enable the rapid evaluation of a much wider set of energies and forces, thus achieving the accuracy of \textit{ab initio} methods while retaining an efficiency similar to that of classical methods. \CC{For example, recently, Ryczko \textit{et al.}\cite{ryczko2022machine} successfully employed Kernel Ridge Regression to predict atomic contributions to the DMC total energy for solid and liquid systems.} The forces necessary for performing molecular dynamics (MD) or geometry relaxation calculations can be predicted by either learning a potential energy surface (PES) and computing its gradients\cite{morawietz2012neural,behler2015constructing} or predicting forces directly from learned force fields.\cite{sauceda2019molecular,chmiela2019sgdml} Conventionally, it is most efficient to learn force fields directly from forces provided to the ML model because the forces can provide richer information regarding the local environment and the curvature of the potential energy surface.\cite{sauceda2019molecular} Nonetheless, because of the challenges that certain QMC methods face, it would be infeasible and extremely expensive to assemble data sets containing forces for training; in this manuscript, we instead focus on the QMC-inspired task of learning potential energy surfaces from which forces can be derived.

Most machine learning techniques used to represent high-dimensional PESs fall into two categories: kernel-based representations\cite{bartok2010gaussian,bartok2015g,rupp2015machine} and deep neural networks.\cite{jose2012construction,behler2007generalized,behler2016perspective,schutt2018schnet} Formally, both kernel- and neural network-based methods can achieve any desired level of accuracy given sufficiently abundant training data, meaning that the choice of method is typically motivated by practical considerations. One neural network approach that has fallen into particular favor because of its simplicity, generality, and efficiency\cite{lee2019simple, khorshidi2016amp,shuaibi2020enabling} is the Behler-Parrinello Neural Network (BPNN) approach, which was first introduced by Behler \textit{et al.}\cite{behler2007generalized} In this approach, the total energy of a system, $E_{tot}$, is represented as the sum of contributions from different types of atoms, $E_{I}$. This quite crucially makes BPNNs size-invariant such that a well-trained BPNN can, in principle, be applied to systems of any size so long as they contain the same elements. Since the total energy produced by BPNNs is a function of the input descriptors, which are in turn functions of the atom positions, forces on the atoms can be taken in a relatively simple fashion within this framework by merely taking the derivative of the total energy with respect to the $x$, $y$, and $z$ coordinates using the chain rule. BPNNs are furthermore convenient because of the many, high-quality packages that have recently been made publicly available for training them.\cite{khorshidi2016amp,lee2019simple, schutt2018schnetpack}

In this work, we thus explore the ability to learn DMC forces from DMC energy data - with no forces provided - by training BPNNs. This has the same flavor as previous work along these lines using Gaussian Processes, but our focus is on more flexible neural networks.\cite{archibald2018gaussian} We test the power of our neural networks by analyzing the accuracy achieved in molecular dynamics simulations and geometry optimizations of three small molecules of increasing complexity, \ch{C2}, \ch{H2O}, and \ch{CH3Cl}. We focus on DMC energies and forces because of the particularly large expense that accompanies the evaluation of DMC forces\cite{tiihonen2021toward} and the fact that it is one of the most popular and well-supported of the QMC methods.\cite{kim2018qmcpack, kent2020qmcpack, wines2020first, staros2022combined} We furthermore focus on small molecule dynamics because of the steep challenge they pose; dynamics can be highly sensitive to inaccuracies in potential energy surface gradients that are not readily exposed by straightforward calculations of MAEs, particularly when multidimensional surfaces are involved. In many situations, the ultimate goal of learning forces is to perform molecular dynamics and optimization simulations, so these serve as practical tests of the ultimate utility of our algorithm. We additionally use this as an opportunity to quantify the effect of forces and the error bars that inherently accompany energies produced by stochastic methods on the models we learn. In so doing, we demonstrate that, while the accuracy with which we can learn force fields plateaus without the knowledge of explicit forces, the force fields learned are sufficiently accurate to produce dynamics that can yield meaningful estimates of bond lengths and energies and can optimize molecular geometries with negligible percent errors - at a fraction of the usual cost. Importantly, the errors on our predictions are smaller than the errors one would obtain using Density Functional Theory to estimate DMC geometries and dynamics, as is common practice. This work thus illustrates the powerful role that machine learning could serve in not only accelerating, but providing key quantities like forces from stochastic and other techniques (e.g., RPA) that do not directly provide them. It also adds to the growing body of work leveraging a range of surrogate methods\cite{tiihonen2022surrogate, chen2022structural, dornheim2021machine} to accelerate and extend the reach of stochastic methods.

This article is organized as follows. First, we briefly introduce DFT and DMC, the \textit{ab initio} methods we use for generating the training dataset, as well as the BPNN method we use for constructing our neural network models. We then present our results. We begin with a discussion of the effects of training without forces and error bars on the accuracy of our force fields. We subsequently demonstrate the performance of our methods on small molecule molecular dynamics and geometry optimization. We finally summarize our findings and address possible future directions.

\section{Computational Methods}\label{sec:methods}

To predict the ``DMC'' dynamics and optimized geometries of our example molecules, we first generate a molecular PES using a neural network (NN) trained on the DMC energies of sampled configurations. Departing from most previous works, we do not train on forces, which can be challenging and/or expensive to obtain from DMC simulations. We then use the trained NN models to supply the forces and energies required by molecular dynamics (MD) or geometry optimization simulations that we subsequently employ to compute time-averaged properties or perform geometry relaxation.

\subsection{Electronic Structure Calculations of Dataset Energies}\label{subsec:estructure}

To train our neural network (AMPtorch) models, sets of DFT and DMC training energies that well-represent the PESs of the systems studied were prepared. DFT calculations were performed using the Atomic Simulation Environment (ASE) package.\cite{larsen2017atomic} PySCF\cite{sun2018pyscf} was used to compute the DFT energies and forces for \ch{H2O} and \ch{CH3Cl}, while QuantumESPRESSO\cite{QE-2017} was used to compute those for \ch{C2}. All DMC calculations were performed using QMCPACK\cite{kim2018qmcpack} and managed using Nexus,\cite{krogel2016nexus} a workflow automation system integrated with the QMCPACK, PySCF, and QuantumESPRESSO packages. The use of Nexus ensured that all DFT outputs could be seamlessly used as inputs to the DMC simulations subsequently performed. 

For the DFT calculations carried out using PySCF, energies and forces were calculated using the Perdew-Burke-Ernzerhof (PBE) functional\cite{perdew1996generalized} with the correlation-consistent, triple zeta basis set (cc-pVTZ). For the DFT calculations carried out using QuantumESPRESSO, the Local Density Approximation (LDA) functional was employed and the Kohn-Sham equations were solved self-consistently within a plane-wave basis set. The kinetic energy cutoff for the DFT wave functions was set to 150 Ry, the value of the Gaussian spreading was set to 0.0002, and the convergence threshold for the energy was set to $10^{-8}$ Hartree. Calculations were performed using the k-points (1 1 1 0 0 0) since only single, isolated molecules were analyzed in this work. Even though one could use larger or more efficient basis sets and more accurate functionals for these calculations, the focus of our study is on the performance of machine learned force fields trained with/without forces, and thus we were less concerned with the absolute accuracy of our energies.

\begin{figure}[]
    \centerline{\resizebox{12cm}{!}{\includegraphics{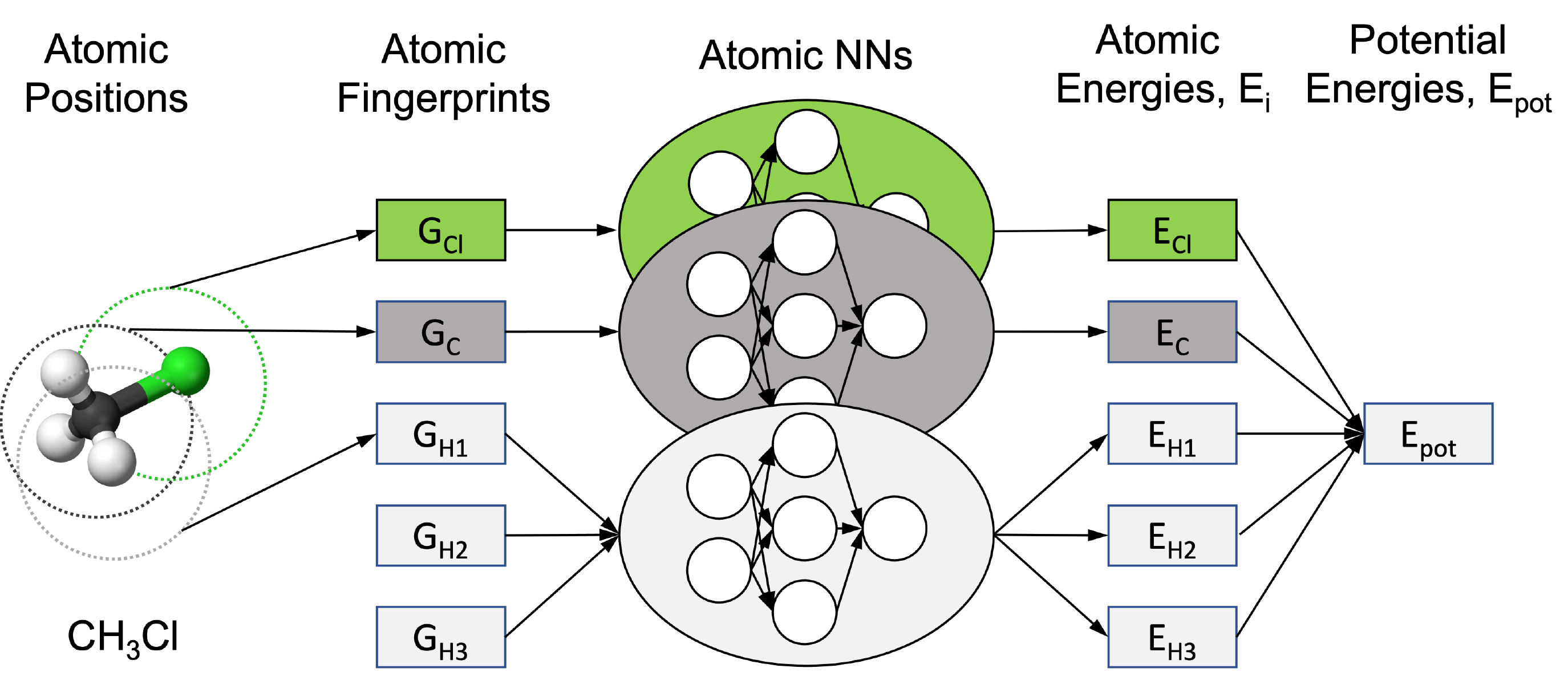}}}
    \caption{A Behler-Parrinello neural network model of \ch{CH3Cl}. For each atom, a feed forward atomic neural network is used to calculate the per-atom energy, $E_i$. The positions of each atom are first transformed into fingerprints consisting of symmetry function vectors, ${\bf G}$. The dashed circles around the atom depict the contributions of different atoms within the cutoff radius to its symmetry functions. Atomic neural networks are then trained on the atomic fingerprints and yield the atomic energies. The total potential energy is obtained by adding all of the atomic energies together.}
    \label{fig:atomicNN}
\end{figure}

For our DMC calculations, we employed the QMCPACK package.\cite{kim2018qmcpack} DMC is a stochastic, many-body method that obtains ground state energies by propagating a trial, typically DFT, wave function in imaginary time. The resulting wave function is represented by an evolving population of walkers that, given sufficient imaginary time, will evolve into the ground state wave function, from which ground state observables can be measured.\cite{foulkes2001quantum} As is conventional in DMC calculations, we first performed DFT calculations to generate a trial wave function that is used as input into Variational Monte Carlo (VMC) calculations. VMC optimizes one-, two-, and three-body Jastrow factors that are then applied to the DFT wave function to introduce dynamical correlation into the trial wave function. To save time, we often reused Jastrow parameters from VMC calculations performed at equilibrium geometries for non-equilibrium geometries. These wave functions were then used as the trial wave function starting points in our DMC calculations. For calculations of the same molecule at different geometries, DMC parameters, including the number of blocks and steps per block, were fixed to ensure consistency (see Table \ref{tab:dmc_parameters}). All calculations were performed with open boundary conditions. Both the DFT and DMC calculations for \ch{C2} used Burkatzki-Filippi-Dolg pseudopotentials,\cite{burkatzki2007energy} whereas correlation consistent Effective Core Potentials (ccECPs) from the QMCPACK pseudopotential library \cite{annaberdiyev2018new, bennett2017new} were used for \ch{H2O} and \ch{CH3Cl}.

\begin{table*}[t]
    \centering
    \begin{tabular}{c|cccc|cccc}
        \hline
        \multicolumn{1}{c|}{\multirow{2}{*}{System}} & \multicolumn{4}{c|}{VMC} & \multicolumn{4}{c}{DMC}\\ 
        \cline{2-9}
        \multicolumn{1}{c|}{} & blocks & steps & timestep & \multicolumn{1}{c|}{samples} & blocks & steps & timestep & walkers \\ \hline
        \ch{C2} & 20 & 10 & 0.4 & 2048 & 400 & 32 & 0.01 & 2048\\
        \ch{H2O} & 200 & 20 & 0.1 & 2048 & 400 & 40 & 0.01 & 2048\\
        \ch{CH3Cl} & 200 & 20 & 0.1 & 2048 & 400 & 40 & 0.01 & 2048 \\
        \hline
    \end{tabular}
    \caption{QMCPACK parameters used in DMC calculations of the molecules studied in this work. Here, the parameter \textit{steps} is the number of step per block used in the DMC calculation.}
    \label{tab:dmc_parameters}
\end{table*}

\subsection{Atom-Centered, Behler-Parinello Neural Networks \label{subsec:networks}}

To learn PESs on which to run our geometry optimization and molecular dynamics simulations, we trained Behler-Parinello Neural Network (BPNNs) on our DFT and DMC energies.\cite{behler2007generalized} While traditional, feed-forward neural networks can be used to construct PESs,\cite{valentini2020constructing} these networks tend not to be practical for learning high-dimensional PESs because they need to be retrained every time new atoms, which are represented by new input nodes, are added to the system. As Behler realized, a more efficient way of learning the potential energy of a system, $E_{pot}$, may be developed by recognizing that the total energy of an $N$-particle system may be decomposed into the sum of $N$ atomic energies, $E_{i}$,\cite{behler2007generalized} 

\begin{equation}
    E_{pot}=\sum_{i} E_{i}.
\end{equation}

Instead of learning the potential energy directly, the atomic energies are learned by training different feed forward neural networks for each of the atom types. Each of the $E_i$ thus depends on each atom $i$'s local chemical environment, which may be represented by different many-body symmetry functions, ${\bf G}$. Each atom's local environment is defined by its neighboring atoms $j$ within a preset cutoff radius, $R_c$, which is set to 6 \r{A} in this work. For example, Figure \ref{fig:atomicNN} represents a schematic Behler-Parrinello neural network for fitting the PES of the \ch{CH3Cl} molecule. Each atom's Cartesian coordinate vector ${\bf R}$ is transformed into the many-body symmetry vector ${\bf G}$, which is also called a fingerprint. This pre-processing is essential for constructing ML potentials as inputs for training should be translationally- and rotationally-invariant, which is not automatically ensured by inputs based on Cartesian coordinates. In this work, we employed two types of symmetry functions: radial functions, $G_i^R$, and angular functions $G_i^A$, which describe the radial and angular environments of atom $i$, respectively. These may be defined as
\begin{equation}
    G_i^R = \sum_{j\neq i}^{\text{all}} e ^{-\eta(R_{ij}-R_s)^2f_c(R_{ij})}
\end{equation}
% and 
\begin{equation}
    \begin{split}
        G_i^A = &2^{1-\zeta} \sum_{j, k \neq i}^{\text{all}} (1 + \lambda \cos \theta_{ijk})^\zeta \\
        &\times e^{-\eta(R_{ij}^2 + R_{ik}^2 + R_{jk}^2)}f_c(R_{ij})f_c(R_{ik})f_c(R_{jk}),
    \end{split}
\end{equation}
where
\begin{equation}
    f_c(R_{ij}) = \begin{cases}
        0.5 \times \left[ \cos(\frac{\pi R_{ij}}{R_c}) + 1 \right], & \text{for } R_{ij} \leq R_c \\
        0 & \text{for } R_{ij} > R_c
    \end{cases}
\end{equation}
is a cutoff function that defines the local environment of each atom within a cutoff radius of $R_c$. $\theta_{ijk}$ is the angle centered at atom $i$ formed by atoms $i$, $j$, and $k$; $\eta$ is the width of the Gaussian functions;  $\lambda$ shifts the cosine function maxima; and $\zeta$ controls the angular resolution (i.e., larger values of $\zeta$ yield symmetry functions with narrower widths). After computing the fingerprints for each atom, the vector ${\bf G}$ is used for training the atomic feed forward neural network to yield the decomposed energy $E_i$. Once the values of the $E_i$ are obtained, the potential energy $E_{pot}$ is calculated by summing over the $E_i$.

Analytic gradients of neural networks, which are central for the evaluation of forces, can be obtained by taking derivatives within the neural network. The force component, $F_{k, \alpha}$, on atom $k$ in direction $\alpha$ can be obtained by applying the chain rule

\begin{equation}
    \begin{aligned}
        F_{k, \alpha} & = -\frac{\partial E}{\partial R_{k, \alpha}}\\
        & = -\sum^N_{i=1}\frac{\partial E_i}{\partial R_{k, \alpha}}\\
        & = -\sum^N_{i=1}\sum^{M_i}_{j=1}\frac{\partial E_i}{\partial G_{i, j}}\frac{\partial G_{i,j}}{\partial R_{k, \alpha}}
    \end{aligned}
    \label{eq:force}
\end{equation}
where $M_i$ is the number of symmetry functions of atom $j$. It should be noted that the architecture for each element, such as carbon or hydrogen, can vary, having its own unique numbers of nodes and layers and its own weights, but for a given element, all atomic neural networks are identical. The use of atomic neural networks enables training performed on smaller molecular systems to be used to make accurate energy predictions in considerably larger molecular systems. 

Over the last few years, many software packages have been released that can train atomic NNs to represent PESs. In this work, we use the AMPtorch package\cite{AMPtorch} developed by the Ulissi Research Group, which is based upon the Atomic Machine-Learning Package (AMP) \cite{khorshidi2016amp} developed by Peterson and co-workers. AMP is designed to represent atom-centered PESs for both periodic and non-periodic systems and is seamlessly integrated into the Atomic Simulation Environment (ASE) package, which means that it can be readily combined with other electronic structure packages. AMPtorch builds upon the foundation of the AMP package, but leverages modern flexible neural network frameworks and the power of GPUs for improved performance. It features a variety of state-of-the-art machine learning methods, including techniques such as online active learning.\cite{AMPtorch} \CC{Details regarding our model construction, training and test sets, and hyperparameter tuning results are included in the Supplementary Materials. In particular, our optimized model employs a fully connected neural network structure consisting of three layers with five nodes in each layer for \ch{C2}, and five layers with ten nodes in each layer for \ch{H2O} and \ch{CH3Cl}. }

\subsection{Construction of Training Datasets\label{subsec:training}}

As further discussed below, the machine-learned geometry optimization and molecular dynamics of three molecules, \ch{C2}, \ch{H2O}, and \ch{CH3Cl}, were studied in this work. These molecules were selected to demonstrate the universal ability of neural networks to obtain reliable forces based upon only potential energy information.

\ch{C2} possesses the simplest PES, which is simply a one-dimensional function of its bond distance. As such, training points for \ch{C2} were chosen within the interval [0.7, 3.5]$\times R_{0}$, where $R_0 = 1.242$ \r{A} is \ch{C2}'s equilibrium bond distance. Points were sampled at 0.05$\times R_{0}$ intervals within the ranges [0.7, 0.90]$\times R_{0}$ and [1.10, 1.50]$\times R_{0}$ as energies in these regions can change dramatically. For bond distances around the equilibrium bond distance, intervals of 0.025$\times R_{0}$ were used since more data was needed to train around the important bond minimum. For bond distances larger than 1.5 $\times R_{0}$, a larger interval of 0.25$\times R_{0}$ was employed. Altogether, 29 points were included in the \ch{C2} training dataset. A denser set of points was selected within the same interval for the test dataset, which consisted of a total of 60 test points. 

The PES of \ch{H2O} is slightly more complex as it is a function of two O-H bond distances and one H-O-H bond angle. The \ch{H2O} training dataset was prepared by grid sampling 20 points along the two O-H bonds and 30 points along the H-O-H bond angle. Points were taken within [0.7, 2.0)$\times R_0$, where $R_0=0.969$ \r{A} is the equilibrium bond distance, for each of the O-H bonds and within [0.3, 1.0)$\times 180^\circ$ for the bond angle, resulting in 12,000 points in total. 2000 points were randomly selected within the same intervals for test purposes. 

Lastly, \ch{CH3Cl}, which possesses a relatively high, nine-dimensional PES, was studied to demonstrate the feasibility of applying our machine learning techniques to many-dimensional potential energy surfaces. Because of this molecule's many-dimensional surface, training points must be more deliberately selected than through straightforward grid sampling, which would require a combinatorially-large number of points to be sampled. Here, we use the 44,820 points used for training in the work by Owens \textit{et al.}\cite{owens2015accurate} In this reference, points were generated using energy-weighted Monte Carlo sampling along nine internal coordinates: the C-Cl bond length $r_0$, with $1.3 \leq r_0 \leq 2.95$\r{A}; three C-H bond lengths $r_1$, $r_2$, and $r_3$ with $0.7 \leq r_i \leq 2.45$ \r{A}; three $\angle$ H$_i$CCl interbond angles $\beta_1$, $\beta_2$, and $\beta_3$ with $65^\circ \leq \beta_i \leq 165^\circ$; and two dihedral angles $\tau_{12}$ and $\tau_{13}$ between adjacent planes containing H$_i$CCl and H$_j$CCl with $55^\circ \leq \tau_{jk} \leq 185^\circ$. A randomly sampled \ch{CH3Cl} test set containing 2000 points was generated within the same intervals.

Once these data sets were assembled, AMPtorch models were built upon them with a learning rate of 0.001 to 0.01 and 2000 epochs, which served as a reasonable compromise between training speed and model quality. During training, the data sets were split into two parts, one for training and one for validation, with the validation/training ratio set to 1:9. After the training process was finished, the trained model was tested on the corresponding test data set to evaluate its performance. Training, validation, and test statistics such as Mean Absolute Errors (MAEs) are reported in the Supplementary Materials. 

\subsection{Geometry Optimization and Molecular Dynamics Simulations}\label{subsec:opt_md}

To verify the quality and robustness of the force fields constructed by AMPtorch using only energies, geometry optimization, and molecular dynamics simulations were performed on the three molecules studied. Fully \textit{ab initio} geometry optimization and dynamics calculations can become undesirably slow because of the need to perform electronic structure calculations at each iteration/time step, which provides an opportunity for machine learning techniques that learn the PESs to accelerate such calculations.

Both the geometry optimization and molecular dynamics simulations were performed using modules implemented in the ASE package.\cite{larsen2017atomic} For the geometry optimizations, the atom calculator was set to either a DFT or AMPtorch calculator, as appropriate. The BFGS method was employed to perform the minimization with the convergence criterion, $f_{max}$, set to 0.05 eV/\r{A} for all test cases. To test the geometry optimization, the molecules were perturbed from their equilibrium geometries and allowed to relax. More specifically, in \ch{C2}, the two carbons were stretched to a distance of 2 \r{A}; in \ch{H2O}, the O-H bond distance was stretched to 1.3 \r{A} and the bond angle was increased to 120$^\circ$; and, in \ch{CH3Cl}, the atoms in the molecule were randomly displaced using the \textit{rattle} function within ASE, with the standard deviation set to 0.1 \r{A}.

Molecular dynamics simulations were performed in both the microcanonical (NVE) and canonical (NVT) ensembles to analyze the influence of stochastic and other errors on simulations with and without thermal fluctuations. For the NVE simulations, the velocity Verlet algorithm was employed, while for the NVT simulations performed at T=300 K, the Langevin algorithm was employed with a friction coefficient of 2.5 $s^{-1}$. For simulations in both ensembles, the initial molecular geometries were optimized using DFT before being fed into the subsequent molecular dynamics simulations. Initializing all MD runs to the same geometries enabled a more direct comparison of the subsequent dynamics. An MD timestep of 0.1 fs was employed in all of the simulations presented, which were run for a total of 2 ps in order to analyze their stability. 

\section{Results and Discussion}\label{sec:results}

\subsection{The Influence of Training without Forces on the Quality of AMPtorch Predictions \label{subsec:forces}}

One of the distinguishing features of this study is our attempt to learn potential energy landscapes \textit{without} the usual aid of force data. A natural question to ask is thus how our model's accuracy - and its requisite training - are affected by the lack of such guiding data. We therefore begin by comparing the impact of force data on the quality of DFT predictions of the PES, since forces can readily be generated from DFT calculations and DFT results are deterministic, which eliminates the potentially confounding impact of stochastic errors.

\begin{figure*}[]
    \begin{subfigure}{.49\textwidth}
        \centering
        \centerline{\resizebox{8cm}{!}{\includegraphics{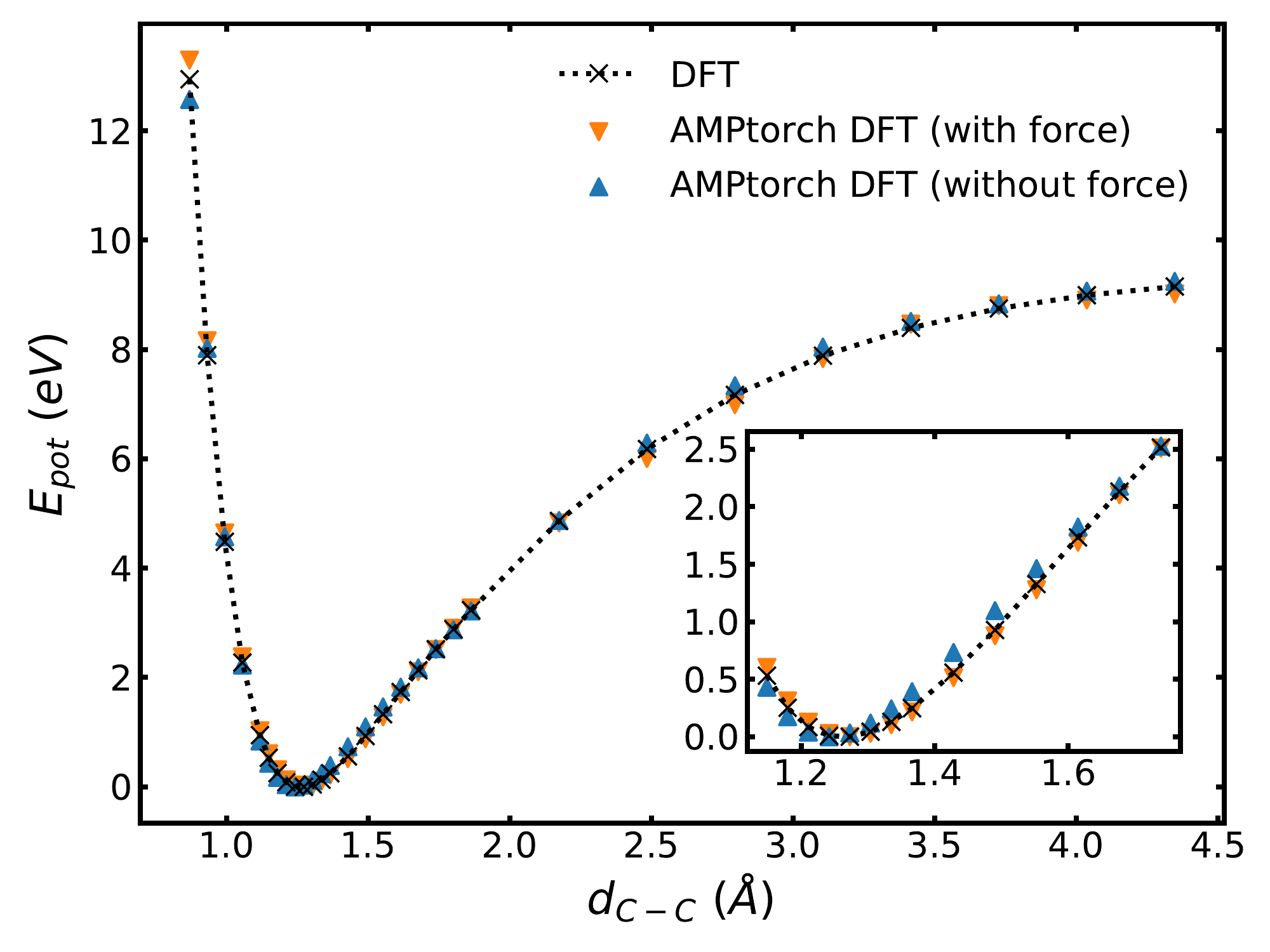}}}
        \caption{}
        \label{fig:c2_dft_energy_comparison}
    \end{subfigure}
    \begin{subfigure}{.49\textwidth}
        \centering
        \centerline{\resizebox{8cm}{!}{\includegraphics{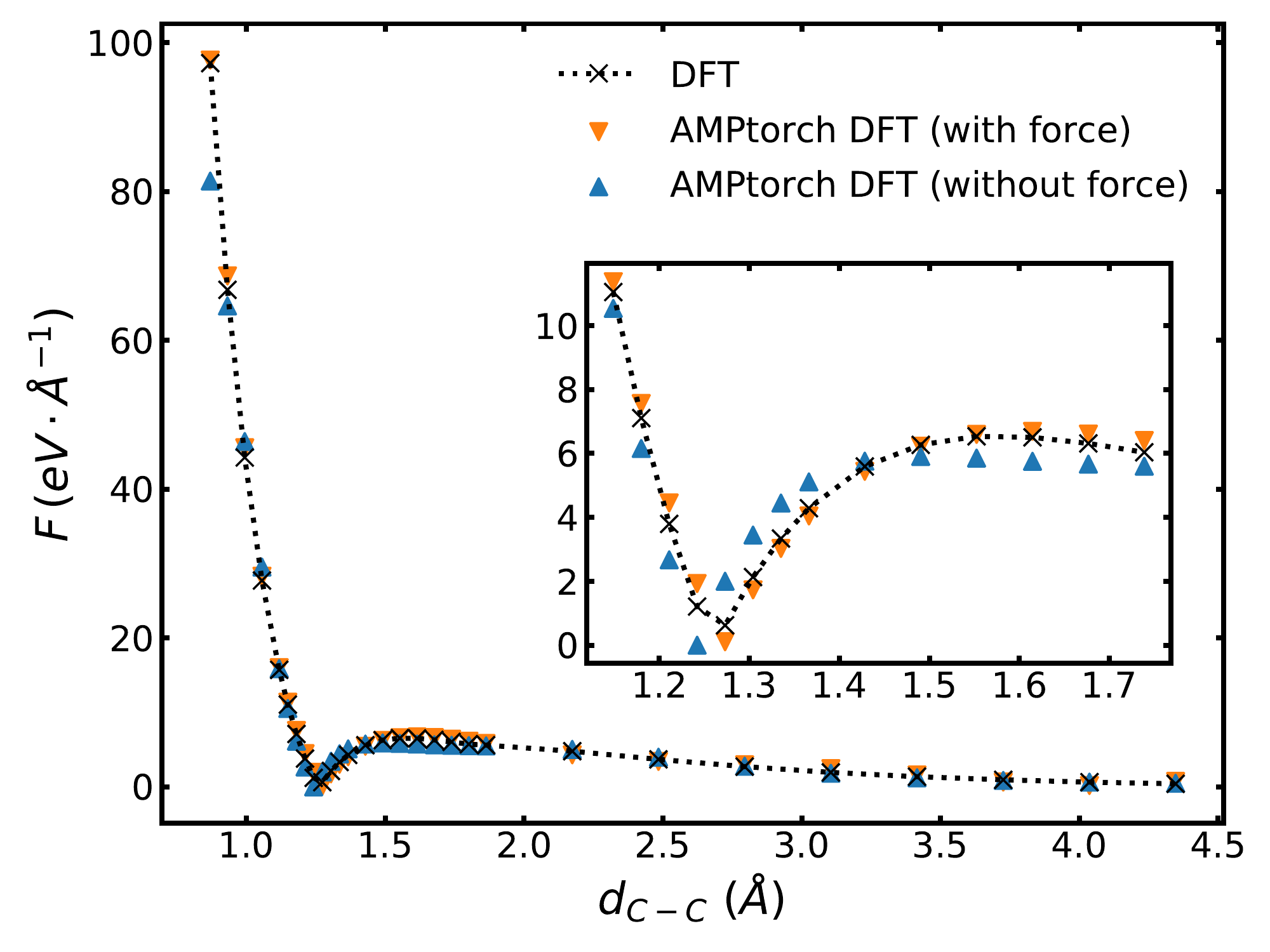}}}
        \caption{}
        \label{fig:c2_dft_force_comparison}
    \end{subfigure}
    \caption{Comparison of predicted (\textbf{Left}) potential energies and (\textbf{Right}) forces along the bond length for \ch{C2} using DFT, an AMPtorch model trained on DFT energies and forces (AMPtorch DFT with force), and an AMPtorch model trained with only DFT energies (AMPtorch DFT without forces). Energies and forces predicted by the AMPtorch DFT model without forces were smaller than the DFT energies at short C-C bond distances, but achieved better agreement at longer distances. In contrast, the AMPtorch DFT model with forces overwhelmingly agrees with the DFT results for both energies and forces across all bond distances.}
\end{figure*}

Figures \ref{fig:c2_dft_energy_comparison} and \ref{fig:c2_dft_force_comparison} show the predictions of the energy of \ch{C2} as a function of the C-C bond distance from two different AMPtorch DFT models, one trained with energies and forces and the other trained with only energies. The predictions from both models have been compared with energies and forces directly calculated by DFT, which can be taken as the ground truth throughout this exploration. As can be observed from the Figure \ref{fig:c2_dft_force_comparison}, the inclusion of force information in the training data set improves the quality of energy and force predictions, particularly at small C-C bond distances, where the first derivative of the potential energy curve is largest. At small bond distances (<1.0 \r{A}), the energies and forces predicted using the AMPtorch model without forces are smaller than those expected based on direct DFT calculations. \CC{For example, DFT calculates the force at 0.87 \r{A} to be 97.25 eV/\r{A}, whereas the AMPtorch model without forces predicts the force at this point to be 81.42 eV/\r{A}}. As one might expect, discrepancies between the calculated and predicted forces are larger than those for the energy because of the lack of force data. This said, the quality of these energy and force predictions dramatically improves at longer bond distances where smaller forces are at play. This is indicative of the poor interpretability of the AMPtorch model for predicting large forces in regions where the magnitudes of forces can change dramatically. Fortunately, for the purposes of modeling the dynamics and relaxation of most molecules, such large force regions are rarely accessed because they are energetically unfavorable. Indeed, the energies at bond distances less than 1.0 \r{A} exceed 2 eV according to DFT calculations, which corresponds to temperatures of greater than $10^5$ K. More specifically, at room temperature, the bond distance between the carbon atoms in the carbon dimer should lie within the interval of [1.22, 1.30] \r{A}. This suggests that, despite this model's large force errors, it and related QMC-based models trained without forces, can still be meaningfully employed in molecular dynamics and other computational tasks that require forces.

This said, it should be noted that the predictions in this region can relatively easily be repaired by adding more points in the desired region or training with knowledge of empirical potentials. The short bond length region of the potential energy curve that describes strong electronic repulsion can be well-described through empirical forms such as the $r^{-12}$ term in the Lennard-Jones potential or the $e^{-Br}$ term in the Buckingham potential. One route recently pursued to facilitate learning bond dissociation curves thus focused on constraining the model to a Lennard-Jones or Buckingham form at these shorter bond distances and enabling the machine learning model to learn the rest of the curve,\cite{mohr2021combining} which we did not find the need to pursue here, but could be readily pursued in other applications.

\begin{figure}[]
    \centering
    \centerline{\resizebox{8cm}{!}{\includegraphics{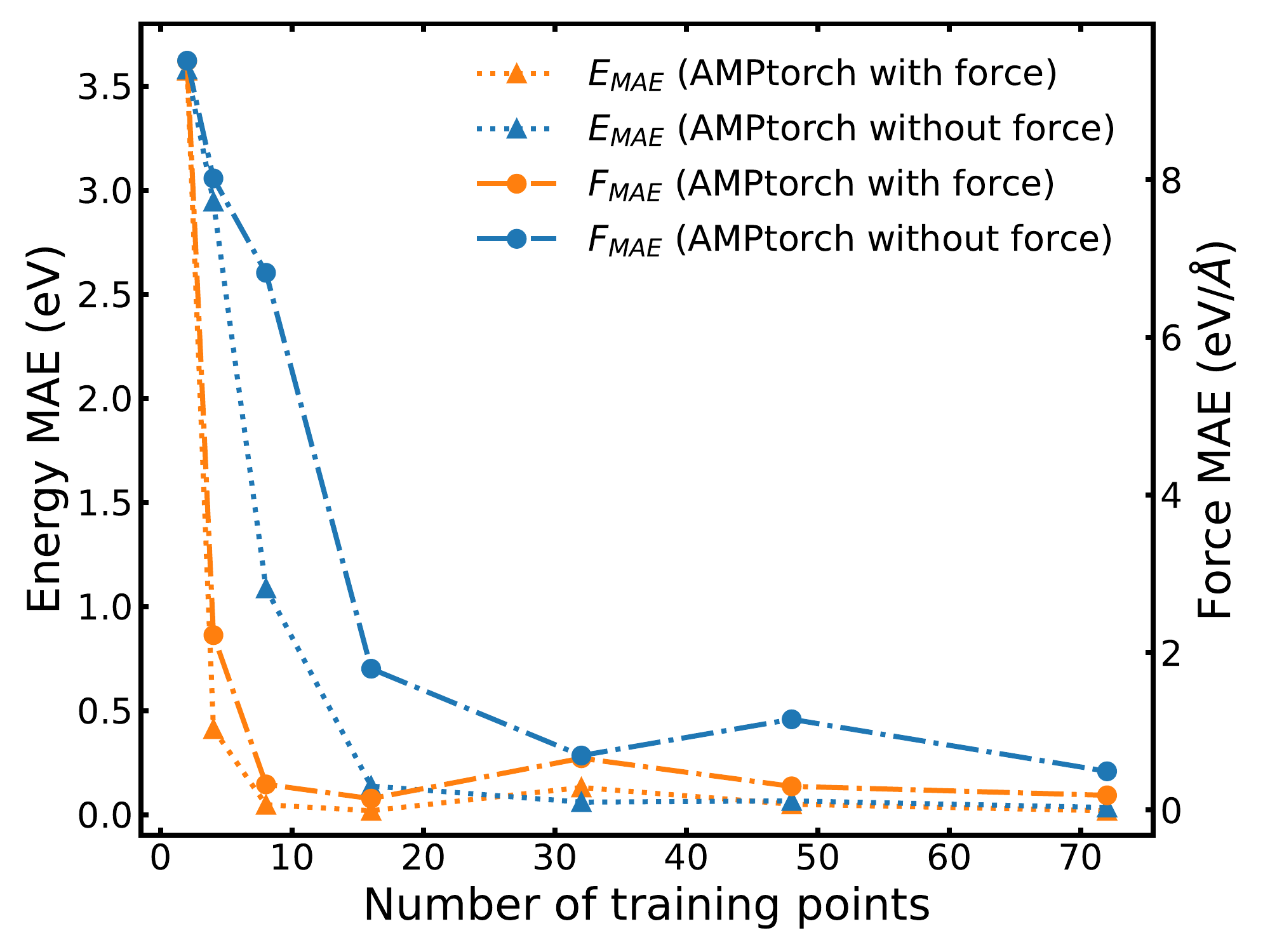}}}
    \caption{Comparison of the performance of the AMPtorch DFT (with forces) and AMPtorch DFT (without forces) models on \ch{C2} trained on an increasing number of points. \CC{For \ch{C2}, each training point consists of 1 energy label and 6 force labels.} The performance is reported by evaluating each model on the test dataset of 48 grid-sampled points.}
    \label{fig:c2_num_points}
\end{figure}

A critical question one may ask to assess the practicality of using energy-only data is \textit{what accuracy can energy-only training achieve relative to training with both energies and forces?} After all, as is known in the computer science literature,\cite{van2018automatic, liao2019differentiable, ma2021comparison} forces provide key information about the slopes of potential energy surfaces that can facilitate prediction. To better understand the consequences of omitting forces, in Figure \ref{fig:c2_num_points}, we compare the energy and force mean absolute errors (MAEs) as a function of the number of training points employed. Starting from an initial set of two points, additional training points are added by grid sampling within the interval [0.7, 3.5]$\times R_{0}$. As expected, the energy and force MAEs decrease with an increasing number of training points until they essentially saturate. The MAEs for the AMPtorch with forces model saturate more quickly than those for the model without forces. \CC{As shown in the Supplementary Materials, this is in large part due to the fact that, when forces are included, there is more data per training point.} Perhaps even more consequentially, while the MAEs on the energies  converge to roughly the same energy for both models, the MAEs on the forces converge to significantly different values depending upon whether forces are employed or not. The force MAE when training without forces saturates to at least a factor of two larger than the force MAE with forces - and this difference does not seem to narrow as the number of training points increases. This implies that models that train with forces will generally do better at subsequently predicting forces, which agrees well with Ref.~\citenum{christensen2020role} and general intuition. Nevertheless, as we will further explore below, the hypothesis is that the errors that can be achieved on the forces, within 0.5 eV/\r{A}, are sufficiently small such that meaningful physics can be obtained from simulations using these forces. 

\begin{figure}[]
    \centering
    \centerline{\resizebox{8cm}{!}{\includegraphics{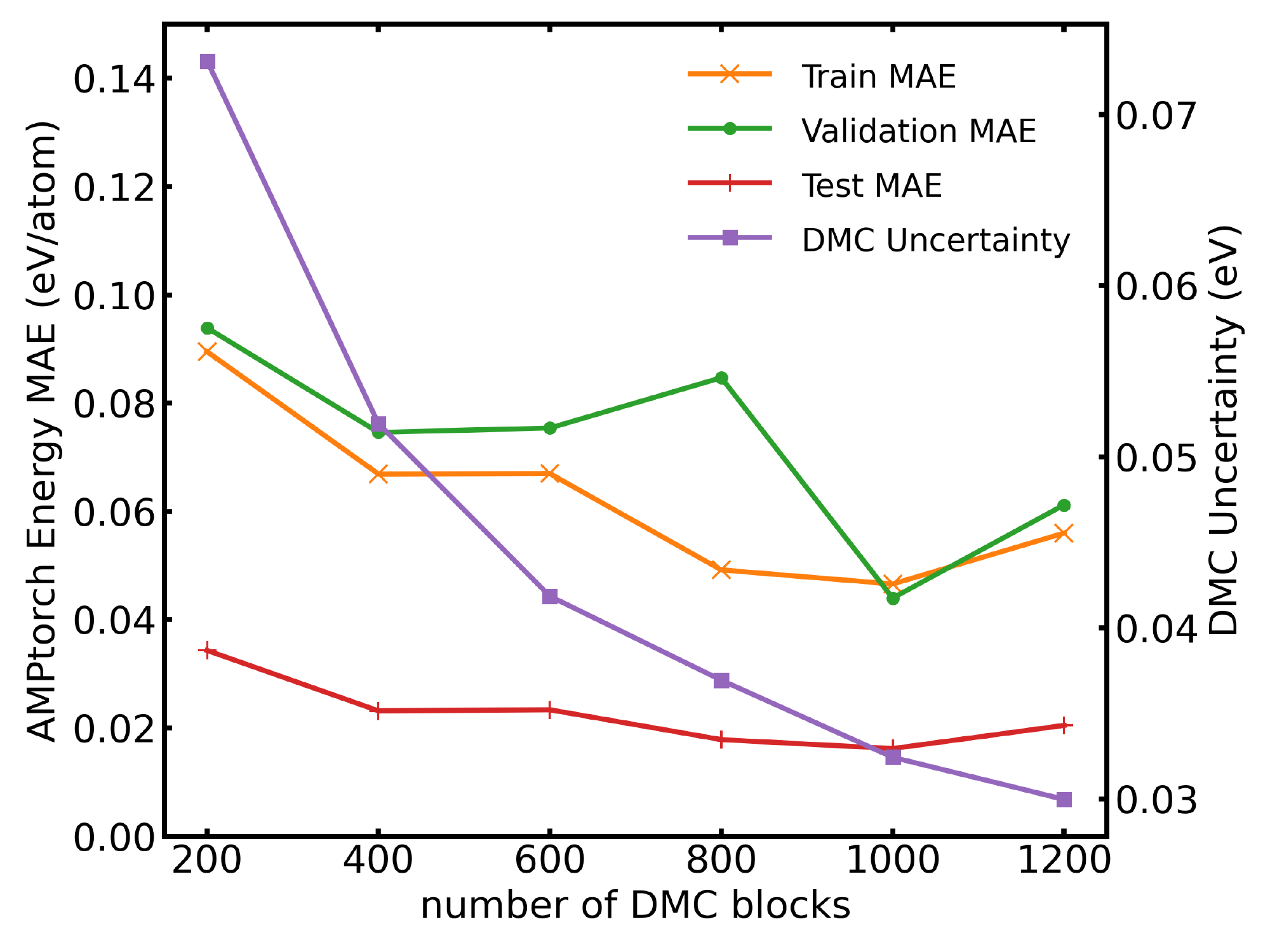}}}
    \caption{The mean average errors (MAEs) vs. the number of DMC blocks for \ch{H2O}. The number of blocks is used as a proxy for the average DMC error bar (DMC Uncertainty). Each block consists of 2048 walkers and the number of DMC blocks is varied between 200 and 1200. As the number of samples increases (i.e., as the DMC uncertainty decreases), the AMPtorch DMC MAEs decrease as well. All MAE values are reported as the best model performance within 2000 epochs.}
    \label{fig:h2o_error_bar}
\end{figure}

\subsection{The Influence of Stochastic Noise on the Quality of AMPtorch Predictions \label{subsec:noise}}

Beyond their lack of forces, QMC methods are also inherently stochastic and thus QMC energies are accompanied by stochastic error bars that can complicate the prediction of reliable forces. During training, the NN-based models employed here attempt to fit the calculated average values of the DMC energies, despite their uncertainties, resulting in PESs with unphysical oscillations and other features, such as derivatives.

To investigate the potential influence of such noise on our AMPtorch energy and force predictions, DMC simulations were performed on \ch{H2O} by varying the number of DMC blocks from 200 to 1200, while keeping all other simulation parameters the same. Average DMC error bars (which we term the `DMC Uncertainty' in Figure \ref{fig:h2o_error_bar}) are expected to decrease with the number of blocks since the magnitude of the stochastic error bars scales as $\frac{1}{\sqrt{n_{samp}}}$, where $n_{samp}$ is the number of samples and the number of blocks is linearly proportional to the number of samples. This is corroborated by the purple line with square dots in Figure \ref{fig:h2o_error_bar}, which decreases with an increasing number of DMC blocks.

As may be anticipated, the training MAE decreases with the number of DMC blocks, i.e., as the DMC error bars are reduced (see the orange line in Figure \ref{fig:h2o_error_bar}). However, reducing the DMC error bars by running simulations longer comes with a steep cost given the overall cost of DMC simulations relative to other techniques. The critical question to ask is therefore how much the DMC MAE, and ultimately, the accuracy of molecular dynamics and geometry optimization calculations, is improved per additional MC block. 

As illustrated in Figure \ref{fig:h2o_error_bar}, there is fortunately a plateau in the DMC MAE that emerges beyond a certain number of DMC blocks. For example, for \ch{H2O}, the DMC MAE does not significantly change beyond 400 DMC blocks. This is because, below 400 blocks, the uncertainty on the DMC energies is sufficiently large that the average energies have not yet converged to their large-sample limit. This causes the neural network to incorrectly learn spurious fluctuations or run into difficulties fitting a smooth PES. Nonetheless, by 400 blocks, the average DMC energies have essentially converged (within 0.05 eV) to their large-sample limit values and spurious fluctuations between points have been averaged away. As a result, by 400 blocks, the AMPtorch model is able to learn smooth fits to the PES. While it is true that sampling for more blocks will further decrease the magnitude of the statistical error bars on these averages, we find that, since the AMPtorch model directly learns the averages, not the error bars, this reduction in error bars matters much less than the convergence of the averages. This explains why the MAE plateaus: beyond the number of blocks at which the average energies converge, the model does not learn any further useful information. Overall, this suggests that, for the purposes of machine learning, DMC error bars only need to be reduced to the point that DMC average energies remain relatively fixed and one does not have to fully converge statistical error bars. This is the reason why we employ 400 DMC blocks in all of our subsequent simulations.

\subsection{Performance of AMPtorch Models for Small Molecule Molecular Dynamics and Geometry Optimization Calculations}

Equipped with an understanding of how our predictions are affected by statistical errors and a lack of forces, we proceeded to analyze how accurately our models can relax the geometries of small molecules and reproduce properties such as average bond distances using molecular dynamics simulations. While many works focus on how well their energy and force predictions can match reference predictions, it is our belief that a true test of the quality of these energies and forces is whether they can be employed for practical purposes in molecular simulations. Moreover, molecular dynamics and relaxation can both, \textit{in principle}, be robust to statistical or other small errors. Indeed, MD simulations are performed at finite temperatures (i.e., in the NVT ensemble) and thus inherently possess their own thermal fluctuations, which could exceed - and therefore mask - fluctuations that arise from errors modeling the PES. Geometry optimization is also typically performed in an iterative fashion that can recover from errors introduced early on in the optimization process. Thus, while comparisons between model and \textit{ab initio} energies or forces serve as the most stringent tests of model accuracy, dynamics-based tests are not only more practical, but may also be more forgiving.

In the following, we therefore perform MD and geometry relaxation using forces learned from QMC energies on small molecules with increasing numbers of internal degrees of freedom, beginning with the carbon dimer with just one internal degree of freedom (the C-C bond distance), proceeding to \ch{H2O} with three degrees of freedom, and lastly ending with \ch{CH3Cl} with nine degrees of freedom. We find that, with sufficiently converged DMC energies, we can faithfully reproduce key properties, such as bond distances and angles, even without direct force information.

\subsubsection{Case I: The Carbon Dimer}\label{sec:carbon_dimer}

We first trained our AMPtorch models on the carbon dimer, \ch{C2}, which represents the simplest test for our approach since it can only stretch in one dimension along its bond distance. AMPtorch models were constructed and trained as described in previous section. This training pipeline resulted in three models: an AMPtorch-DFT model with forces, which is produced using AMPtorch trained on both DFT energy and force data; an AMPtorch-DFT model without forces, which is produced using AMPtorch trained on only DFT energy data; and an AMPtorch-DMC model, which is produced using AMPtorch trained on only DMC energy data. We included the AMPtorch-DFT model with forces because it enables us to analyze the impact of including forces on the accuracy of the dynamics and relaxations performed. 

\begin{figure}[]
    \centering
    \centerline{\resizebox{8cm}{!}{\includegraphics{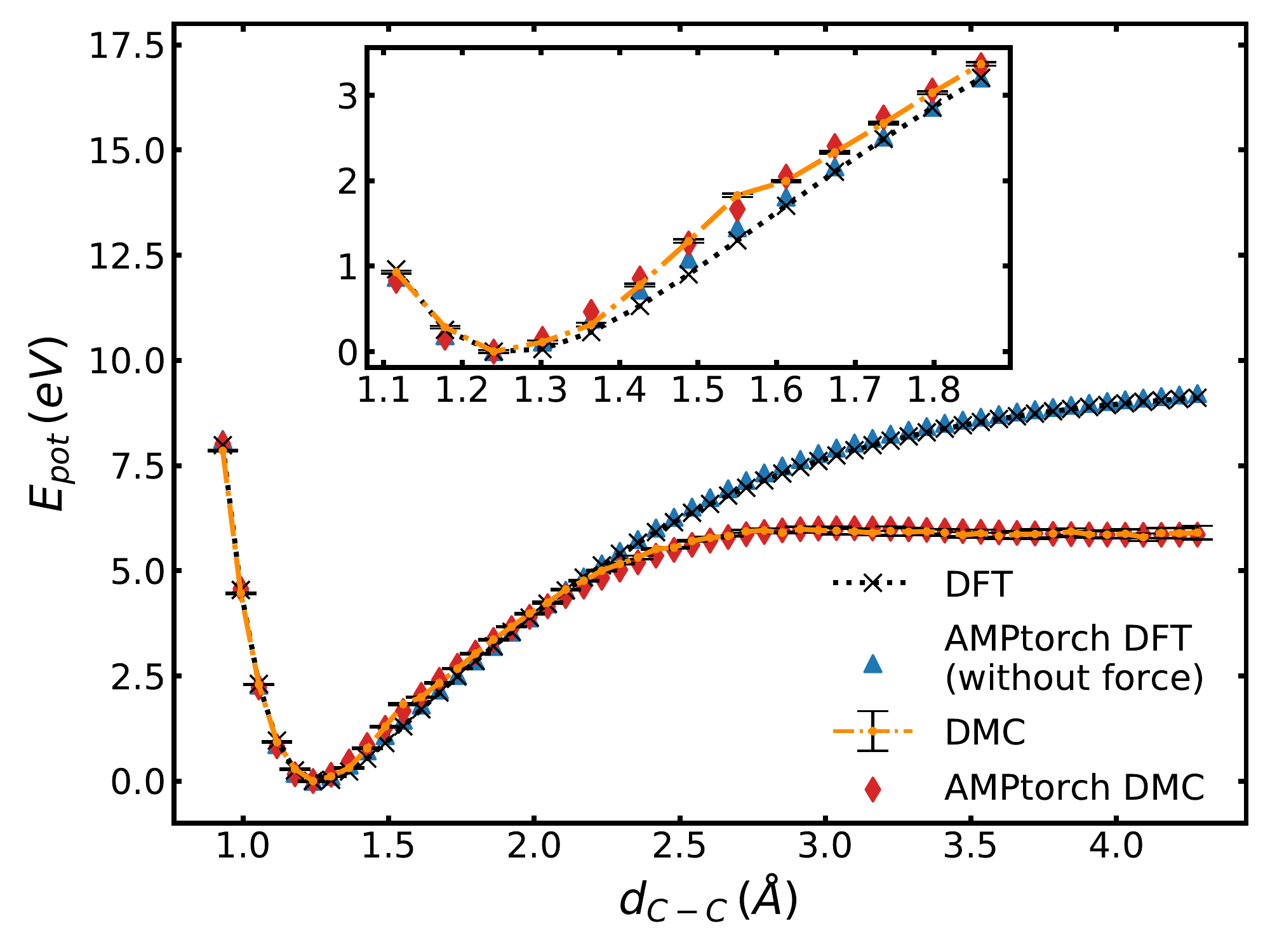}}}
    \caption{Comparison of potential energy curves for \ch{C2} using DFT (black dot line), DMC (yellow dashdot line), an AMPtorch model trained on DFT energies only (blue triangles), and an AMPtorch model trained on DMC energies only (red diamonds). The test data points shown in the figure are evenly distributed in the range from 0.93 to 4.28 \r{A}. The inset highlights the energies near the bottom of the well.}
    \label{fig:c2_pes}
  \end{figure}

The resulting potential energy curves calculated with DFT, DMC, AMPtorch-DFT (without forces), and AMPtorch-DMC are shown in Figure \ref{fig:c2_pes}. Comparing all four curves, we see that the AMPtorch models are able to accurately reproduce the energies near the equilibrium distance. Some slight deviations (of the order of \CC{0.16 eV} relative to the 0.5 eV difference between the DMC and DFT calculations) between the DMC and AMPtorch DMC curves emerge between 1.5 \r{A} and 1.7 \r{A}. These arise because the DMC points are not smooth in this range, illustrating how the AMPtorch DMC model attempts, but correctly fails, to accommodate such spurious deviations into its fits.
With these deviations included, the test MAE for the AMPtorch-DMC energy is \CC{0.0368 eV/atom}, while that for the AMPtorch-DFT (without forces) energy is \CC{0.0274 eV/atom} (see Table S3). Such small MAE values are noteworthy given the fact that it takes just seconds to compute each AMPtorch-DMC data point compared to the minutes per point it takes to perform full DMC calculations. The corresponding comparison of the force predictions may be found in the Supplementary Materials.

What can also be observed from this plot is that the DFT and DMC calculations differ substantially from each other at long bond distances ($>$2.5 \r{A}). As one may anticipate, because DMC is an inherently multireference method, it can more accurately reproduce the C-C bond dissociation barrier, which DFT estimates to be significantly larger than it actually is. The inset of the Figure \ref{fig:c2_pes} furthermore illustrates that DFT predicts a broader energy well that rises more gradually with increasing bond distance than DMC, which agrees with the trends between DFT results from Ref.~\citenum{dinh2010self} and CCSD result from Ref.~\citenum{bhattacharjee2020comprehending}. We shall see that DMC's steeper well naturally leads to smaller amplitude C-C vibrations in our molecular dynamics simulations.

\paragraph*{Molecular Dynamics Simulations}
As a further practical test of our AMPtorch-DMC model, we next used it to perform MD simulations in the NVE and NVT ensembles. For both ensembles, we initialized the molecules to the equilibrium geometries predicted by their models. We then sampled momenta to apply to each of the atoms according to the Maxwell-Boltzmann distribution at our MD temperature of 300 K and let the dynamics progress. 

Because MD simulations rely upon forces, they are sensitive to any unphysical fluctuations in the potential energy surface or its derivatives. Indeed, even slight aberrations in the PES's derivatives can cause the molecules to first undergo unusual motions and then eventually ``explode" into fragments, as we observed when our models were inadequately trained. Nevertheless, with MAEs on the energies and forces as small as those used to produce Figure \ref{fig:c2_pes}, we were able to stabilize machine learning-informed NVE dynamics based upon QMC energies without forces. 

\begin{figure*}[]
  \begin{subfigure}{.49\textwidth}
      \centering
      \centerline{\resizebox{8cm}{!}{\includegraphics{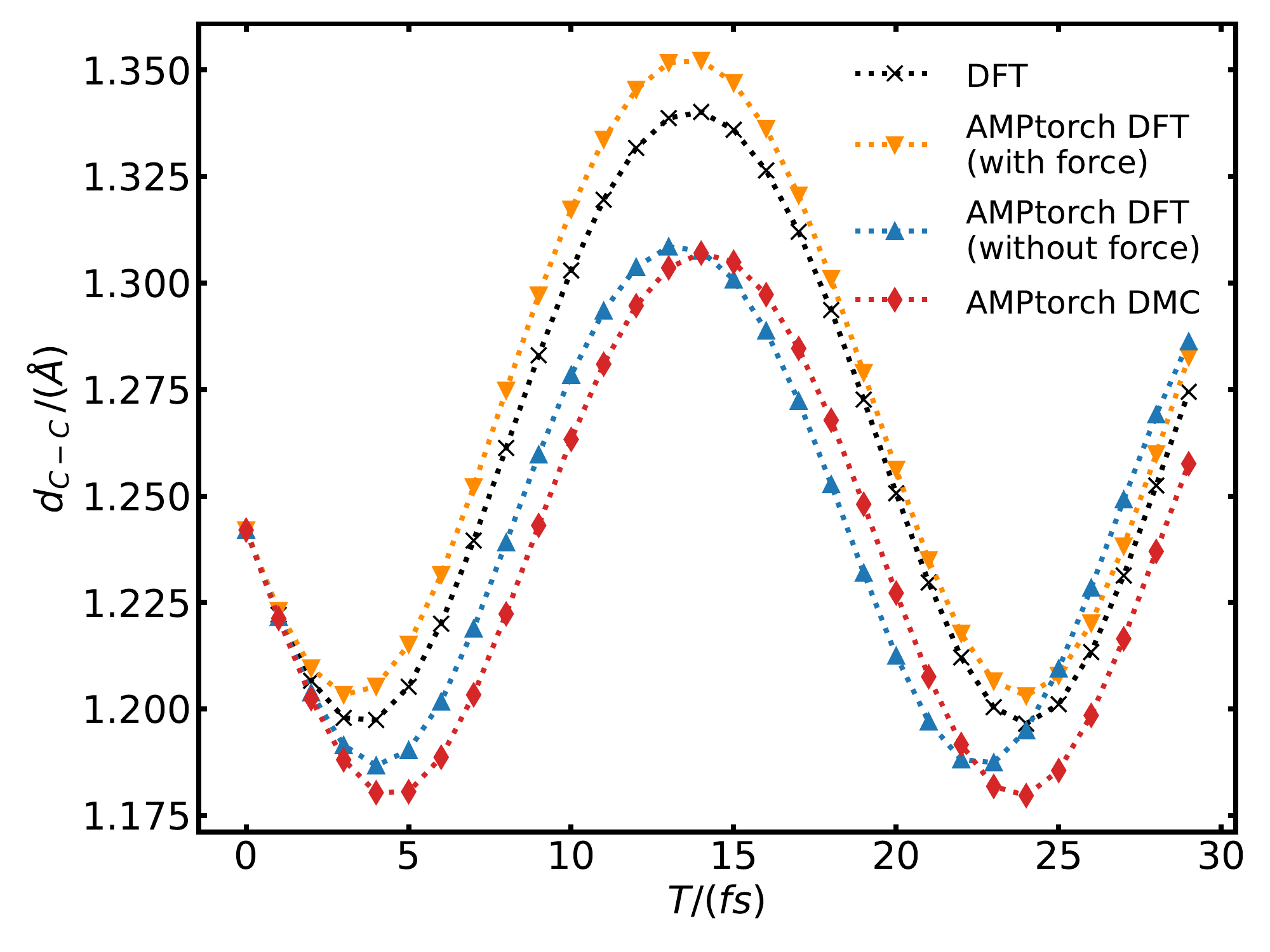}}}
      \caption{}
      \label{fig:c2_nve_r}
  \end{subfigure}
  \begin{subfigure}{.49\textwidth}
      \centering
      \centerline{\resizebox{8cm}{!}{\includegraphics{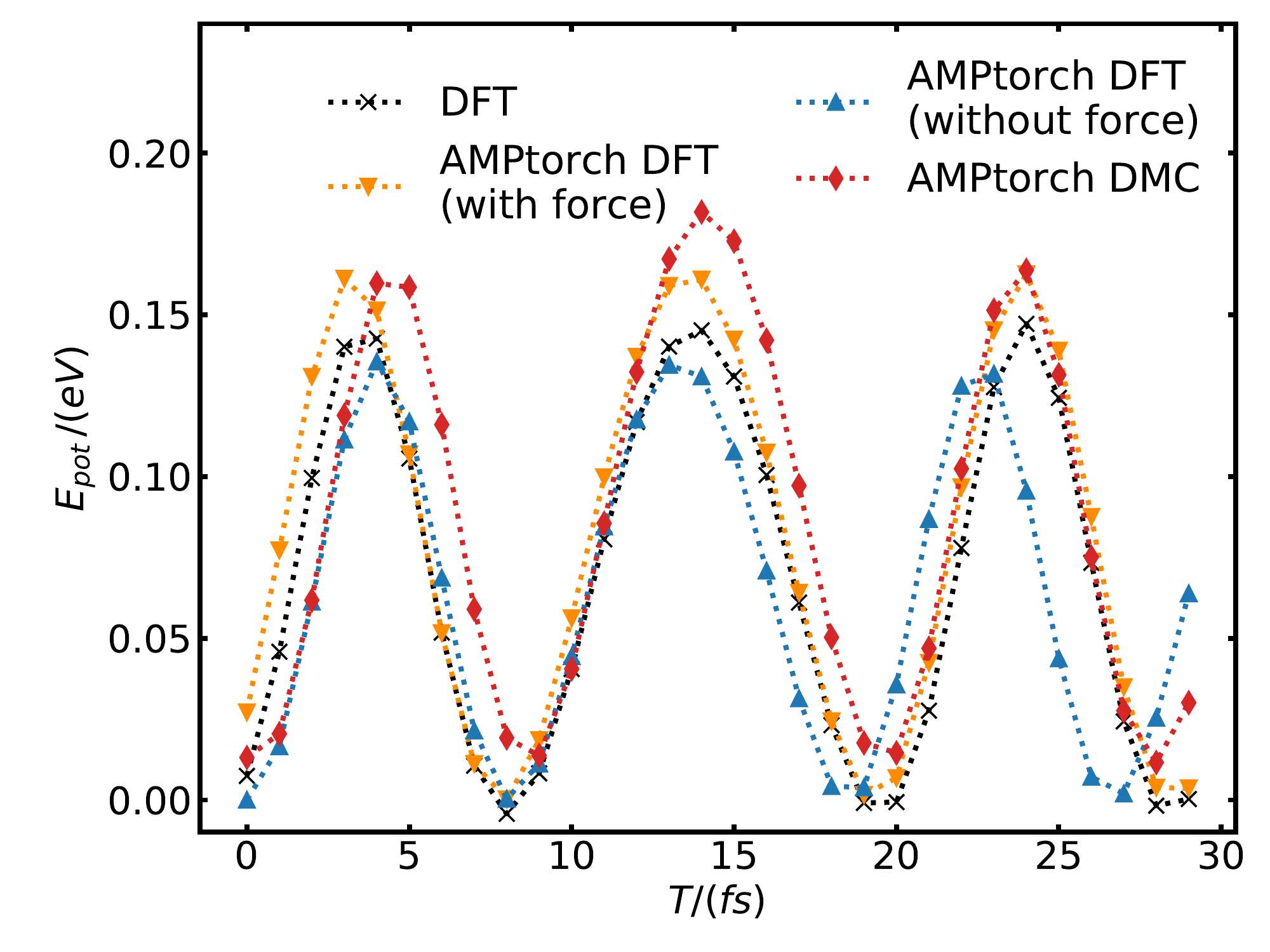}}}
      \caption{}
      \label{fig:c2_nvt_epot}
  \end{subfigure}
  \caption{NVE molecular dynamics simulations for \ch{C2} performed using DFT (black crosses),  an AMPtorch model trained with both DFT energies and forces (orange triangles), an AMPtorch model trained on DFT energies only (blue triangles), and an AMPtorch model trained with DMC energies only (red diamonds). \textbf{(a)}: C-C bond distance vs. time and \textbf{(b)}: E$_{\text{pot}}$ vs. time. The E$_{\text{pot}}$ curves were obtained by subtracting the minimum trajectory energy from the energy at the given time at each point in the trajectory. Here, we only show 30 fs of dynamics for illustration, but the full simulation extends beyond 2 ps.}
\end{figure*}

Despite lacking explicit forces, our \ch{C2} AMPtorch models were able to stabilize more than 2 ps of dynamics. A 30 fs window of this longer trajectory is depicted in Figure \ref{fig:c2_nve_r} for illustration purposes, but also because the fully \textit{ab initio} DFT-based dynamics are expensive and were thus run for only a short duration. For all models, a timestep of 0.1 fs was chosen. This relatively small timestep was selected to minimize the influence of aberrations in the AMPtorch energy surface on the dynamics. This means that forces were updated more frequently than usual, but the cost involved with such small timesteps is readily compensated by the faster calculations made by the AMPtorch models. Based on our simulations, the AMPtorch models can calculate forces over 100 times faster than fully \textit{ab initio} simulations.

Both the DFT and DMC models trained without forces exhibited smooth oscillations in the C-C bond distance over time, however the oscillations exhibited by the AMPtorch DMC simulations were shifted to smaller C-C bond distances. This is because the DMC potential energy surface is steeper than the DFT surface, as can be gleaned from the inset of Figure \ref{fig:c2_pes}. 
A back-of-the-envelope calculation of what the minimal and maximal NVE C-C bond lengths should be based on the DMC potential energy surface given an initial kinetic energy of 0.1356 eV (which is the initial kinetic energy in both the NVE and NVT simulations) suggests that the C-C bond distance should fluctuate between 1.18 \r{A} and 1.30 \r{A}, as can indeed be observed in Figure \ref{fig:c2_pes}.

Since a point-by-point comparison of molecular dynamics simulations does not make sense given the inherent chaos that emerges in numerical MD implementations, we also computed the average C-C bond distances over the course of the dynamics, as provided in Table \ref{tab:c2_md}. The DFT and AMPtorch DFT with forces model bond distances differ by \CC{0.7\% in the NVE ensemble and 0.7\% in the NVT ensemble, while the DFT and AMPtorch DFT without forces model bond distances differ by 1.9\% and 1.2\%}, respectively. These percent differences are smaller than the percent differences between DFT and higher accuracy DMC or other quantum chemistry bond distance predictions, which can be as large as several percent,\cite{gillan2012assessing} meaning that the errors we observe are smaller than the errors that arise from using less accurate electronic structure methods. This signifies that AMPtorch models trained with energy-only data can still be employed to obtain meaningful MD data, which supports our initial hypothesis. Given that DMC should yield different bond distances than DFT, it is unfair to directly compare the DFT and AMPtorch DMC bond distances. Nonetheless, the average C-C bond distance using the AMPtorch DMC model is \CC{1.2450 \r{A} based on NVE dynamics and 1.2354 \r{A} based on NVT dynamics}, which compare favorably with the experimentally-determined bond distance of 1.2425 \r{A}.\cite{booth2011breaking} See Supplementary Material for a further presentation of the full trajectories and bond distance distributions.  

\begin{table}[]
  \centering
  \begin{tabular}{ccccc}
      \hline
      Observables& DFT & \begin{tabular}[c]{@{}c@{}}AMPtorch DFT\\ (with forces)\end{tabular} & \begin{tabular}[c]{@{}c@{}}AMPtorch DFT\\ (without forces)\end{tabular} &
      \begin{tabular}[c]{@{}c@{}}AMPtorch\\ DMC\end{tabular} \\
      \hline
      NVE $d_{C-C}$ (\r{A}) & 1.2720 & \CC{1.2807} & \CC{1.2477} & \CC{1.2450} \\
      NVT $d_{C-C}$ (\r{A}) & 1.2589 & \CC{1.2682}  & \CC{1.2402} & \CC{1.2354} \\
      \hline
  \end{tabular}
  \caption{Mean C-C bond distances in the \ch{C2} molecule from NVE and NVT molecular dynamics simulations using DFT, AMPtorch-DFT (with forces), AMPtorch-DFT (without forces), and AMPtorch-DMC models.}
  \label{tab:c2_md}
\end{table}

NVT simulations at a temperature of T=300 K were similarly performed on the model potential energy surfaces. Their trajectories are presented in the Supplementary Materials and their average bond distances are again presented in Table \ref{tab:c2_md}. These NVT average bond distances are shorter than the NVE bond distances because the average kinetic energy available to the molecule starting from the same initial configuration is smaller at T=300 K than at the energy used in the NVE ensemble calculations. One hypothesis of this work is that, because of their inherent thermal fluctuations, NVT molecular dynamics simulations will be relatively resistant to errors in the learned energies and forces. Indeed, at T=300 K, thermal fluctuations lead to fluctuations in the \ch{C2} energy of around 0.053 eV, which could in turn produce bond fluctuations of up to 0.09 \r{A} based on the \ch{C2} potential energy surface. Altogether, this means that force fluctuations of up to 0.59 eV/\r{A} may be observed for the \ch{C2} molecule at this temperature. In contrast, the MAE on the forces for the AMPtorch DFT model with forces is \CC{0.0598 eV/\r{A} and for the AMPtorch DFT model without forces is 0.0580 eV/\r{A}} (see Table S3) - both of which are smaller than the maximum thermal fluctuations expected to be observed in the simulations. This corroborates our point that NVT MD simulations should be relatively insensitive to reasonable fluctuations in machine learned force fields. \CC{This hypothesis is further corroborated by our \ch{H2O} and \ch{CH3Cl} results.}

\paragraph*{Geometry Optimization}
Another common, practical use of \textit{ab initio} forces is for relaxing molecules from arbitrary or trial geometries to their fully-relaxed, equilibrium geometries for subsequent calculations of their electronic and other properties. We therefore also analyzed the performance of our AMPtorch DFT and DMC models on the structural optimization of \ch{C2}. In these tests, the C-C bond distance was first initialized to 2 \r{A} and the BFGS algorithm was used for optimization, while the AMPtorch models served as the calculators for the forces and energies. The convergence criterion for the optimization was set such that the forces on each atom had to be less than 0.05 eV/\r{A}. Since \ch{C2} is one-dimensional, one would expect the optimization to relax the \ch{C2} bond to its shorter equilibrium length in a relatively small number of iterations. 

\begin{figure}[]
  \centerline{\resizebox{8cm}{!}{\includegraphics{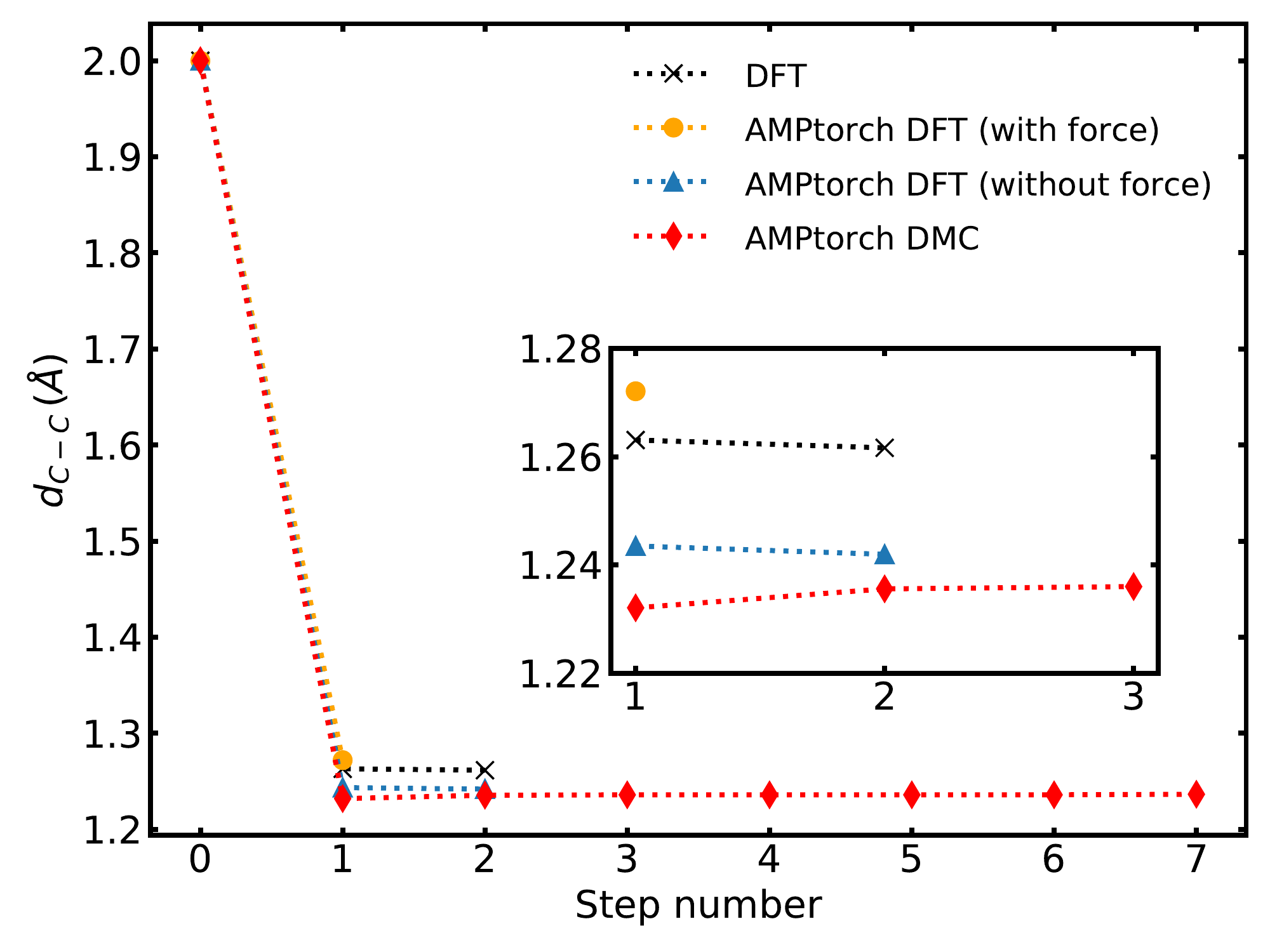}}}
  \caption{\ch{C2} geometry relaxation using DFT and the trained AMPtorch models. The molecular bond length is initialized to 2 \r{A} and DFT/model forces are used to relax it over multiple optimization iterations.}
  \label{fig:c2_bpnn_optimization}
\end{figure}

\begin{table}[t]
  \centering
  \begin{tabular}{ccccc}
      \hline
      \begin{tabular}[c]{@{}c@{}}Step\\ Number\end{tabular} & DFT   & \begin{tabular}[c]{@{}c@{}}AMPtorch DFT\\ (with forces)\end{tabular} & \begin{tabular}[c]{@{}c@{}}AMPtorch DFT\\ (without forces)\end{tabular} & AMPtorch DMC \\
      \hline
      0 & 2.0 & 2.0 & 2.0 & 2.0 \\
      1 & 1.263 & \CC{1.272} & \CC{1.244} & \CC{1.231}\\
      2 & 1.262 & - & \CC{1.242} & \CC{1.236} \\
      3 - 7 & - & - & - & \CC{1.236} \\
      8 & - & - & - & \CC{1.237} \\
      \hline
  \end{tabular}
  \caption{Comparison of the \ch{C2} structural optimization data from simulations using DFT and the trained AMPtorch models: AMPtorch DFT (with forces), AMPtorch DFT (without forces), and AMPtorch DMC. In all cases, the \ch{C2} bond length is initialized to 2 \r{A} and the BFGS method is used for geometry optimization with an fmax value set to 0.05 eV/\r{A}. The reported experimental value for the C-C equilibrium bond distance is 1.242 \r{A}.}
  \label{tab:c2_geom_opt}
\end{table}

A plot of the C-C bond length vs. optimization step number is depicted in Figure \ref{fig:c2_bpnn_optimization} for the fully \textit{ab initio} DFT and our AMPtorch models. \CC{We see from both Figure \ref{fig:c2_bpnn_optimization} and Table \ref{tab:c2_geom_opt} that, even without forces, the AMPtorch DFT without forces and DMC models were able to relax \ch{C2} in just two iterations, which is the same number of iterations required by DFT. The difference between the DFT and AMPtorch DFT (without forces) bond lengths is found to be 0.02 \r{A} (1.6\% error), while that between the AMPtorch DMC result and the experimental bond length of 1.242 \r{A} is 0.005 \r{A}.} The AMPtorch DMC optimizations are thus accurate to within the few percent accuracy needed for most subsequent electronic structure calculations - and more accurate than all of the other methods studied here. \CC{Interestingly, the AMPTorch DFT (with forces) models relax the dimer in only one iteration to a bond length that is 0.01 \r{A} (0.8\%) different than the DFT length.} This suggests that the AMPtorch model leads to slightly different bond lengths than DFT regardless of the inclusion of forces in the training dataset. From this data, the cost for not incorporating forces into the AMPtorch DFT without forces or AMPtorch DMC models is the need for additional iterations. Even though both of these forceless models near convergence within one iteration, the absence of forces seems to cause the bond lengths to fluctuate slightly more around their minima, \CC{which is most evident in the AMPtorch DMC model,} frustrating the satisfaction of the convergence criterion. The extra iterations required do not come with additional computational costs: the AMPtorch models are found to perform this optimization $10^2$ faster than fully \textit{ab initio} DFT. This strongly suggests that optimization calculations, which often require repeated calls to energy and force routines, may be fertile ground for acceleration by machine learning techniques, even in the absence of force data.

\subsubsection{Case 2: Water Molecule}\label{sec:water}

Given the success of our machine learning models on the \ch{C2} dimer, which can be viewed as a relatively simple case because of its one-dimensional potential energy curve, we next considered how our model performs on \ch{H2O}, which now has three internal coordinates: two O-H bond distances and one H-O-H bond angle. While the two O-H bonds should be symmetric, we find that, if we do not sample the O-H bonds separately, the dynamics will not properly sample both O-H bond lengths, heavily favoring the stretching of one over the other. The presence of a bond angle also adds an additional subtlety to our modeling because the well-depths for the potential energy as a function of such angles are often relatively shallow compared to the well-depths for bond lengths.

As before, we began by training our models on DFT and DMC energies sampled using grid sampling, as described in training datasets construction section. The presence of multiple bonds and angles in our training serves as a further test of our grid-based sampling methodology because it must now sample enough points along multiple inequivalent dimensions. The training and test MAEs are reported in \CC{Table S4}. As we can see, AMPtorch models trained with energy-only data are able to make competitive energy predictions, but struggle to predict forces, with force MAE values \CC{two times} larger than the force MAE values from the AMPtorch model trained with forces data. This implies that, as the dimension increases, the PES possesses a larger number of non-trivial features that are more difficult to learn through the straightforward addition of training data.

\begin{figure*}[]
    \begin{subfigure}{.49\textwidth}
        \centerline{\resizebox{8cm}{!}{\includegraphics{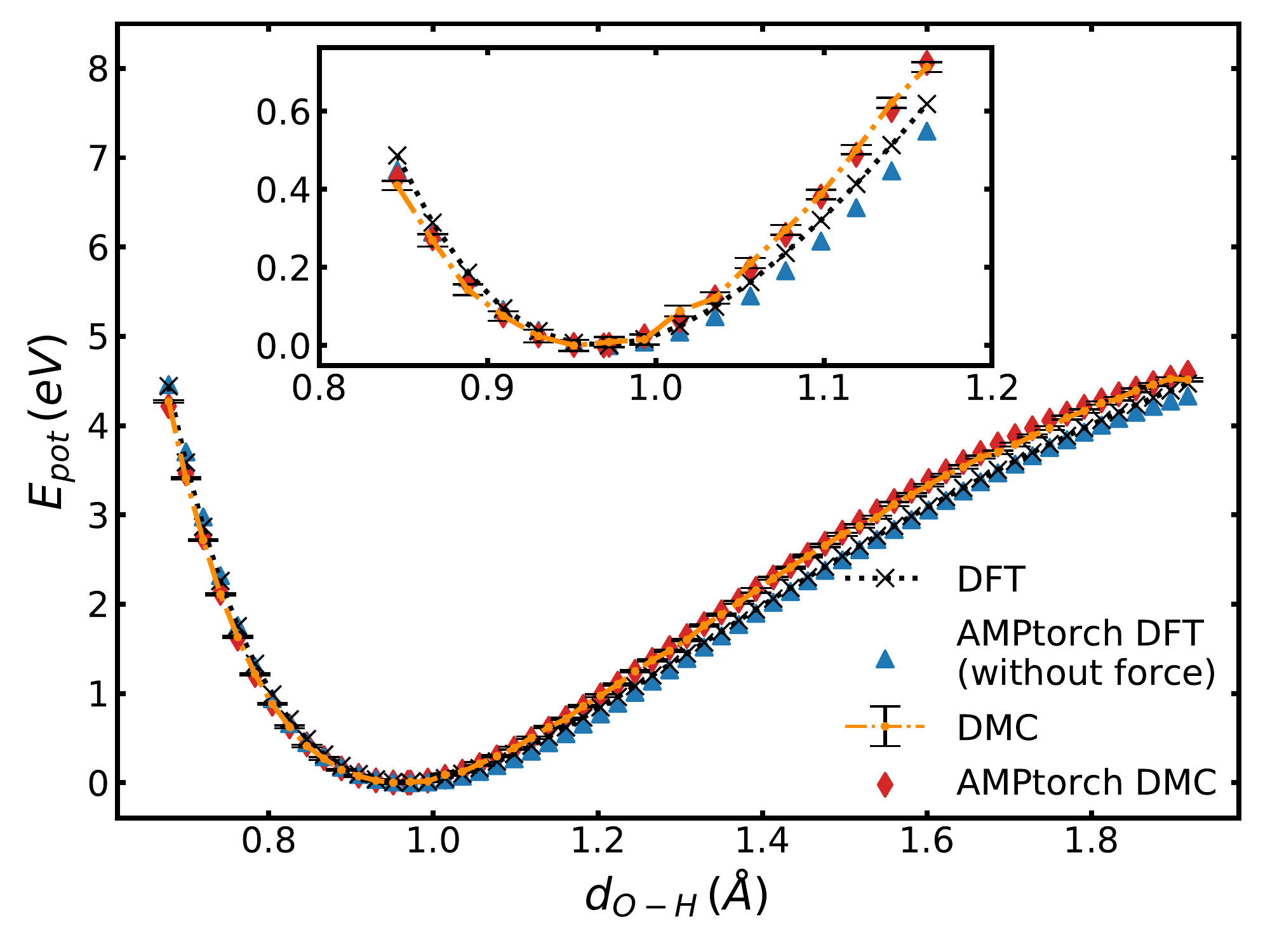}}}
        \caption{}
        \label{fig:h2o_validation_R}
    \end{subfigure}
    \begin{subfigure}{.49\textwidth}
        \centerline{\resizebox{8cm}{!}{\includegraphics{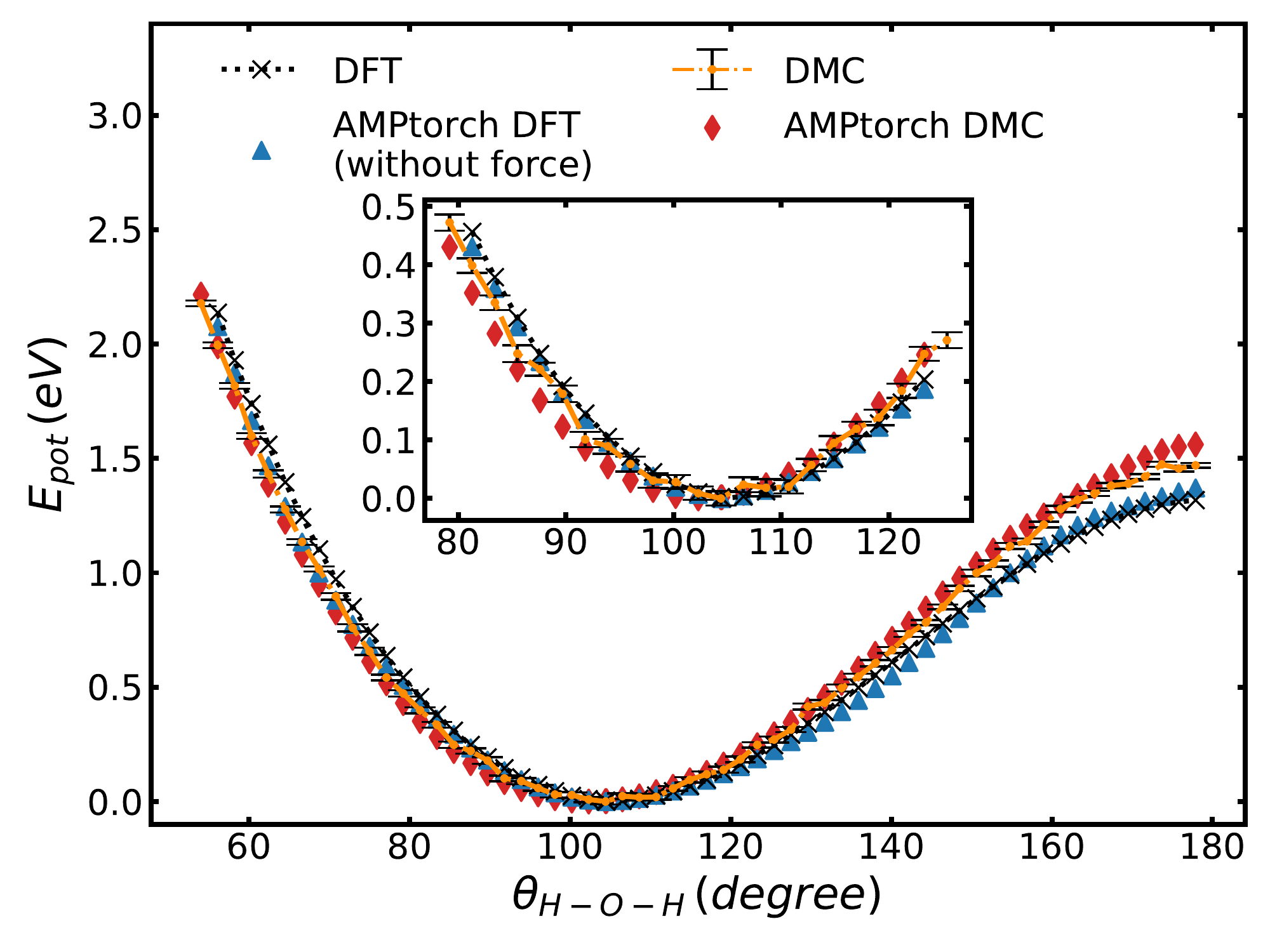}}}
        \caption{}
        \label{fig:h2o_validation_theta}
    \end{subfigure}
    \caption{
        A comparison of the potential energy curves for the \ch{H2O} bond distance, $d_{O-H}$, and bond angle, $\theta_{H-O-H}$, produced using DFT, DMC, AMPtorch-DFT, and AMPtorch-DMC. \textbf{(a)}:
        E$_{\text{pot}}$ vs. the O-H bond distance, where the bond angle is set to the equilibrium bond angle, 104.5$^\circ$;  \textbf{(b)}: E$_{\text{pot}}$ vs. the H-O-H bond angle, where the bond distances are set to the equilibrium bond distance, 0.969 \r{A}.}
  \end{figure*}

Moreover, comparisons between the \textit{ab initio} DFT and AMPtorch model results are presented in Figures \ref{fig:h2o_validation_R} and \ref{fig:h2o_validation_theta}, which show how the energy is predicted to vary as a function of the O-H bond distance and the H-O-H bond angle. Despite the presence of noise in the DMC calculations, the AMPtorch DMC model successfully agrees with the DMC points in most regions. All of the simulation techniques employed yield roughly the same potential energy curves as a function of bond distance, with the DMC curves being slightly steeper at longer bond distances (see the inset in Figure \ref{fig:h2o_validation_R}), similar to what was also seen for the \ch{C2} dimer. \CC{Along the bond angle dimension, all of the models predict similar energies except that the DMC model predicts the minimum to occur at slightly smaller bond angles and predicts larger energies at the largest bond angles.}

\paragraph*{Molecular Dynamics Simulations}

\begin{figure*}
\centering
\begin{subfigure}[b]{\textwidth}
    \centering
    \includegraphics[width=\textwidth]{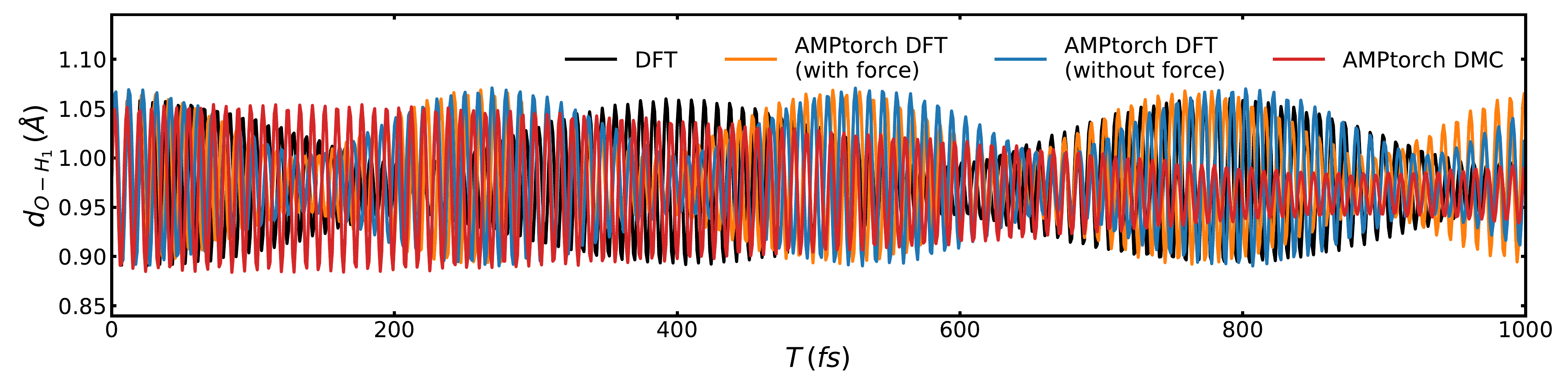}
    \label{fig:h2o_nve_oh1}
\end{subfigure}
\hfill
\begin{subfigure}[b]{\textwidth}
    \centering
    \includegraphics[width=\textwidth]{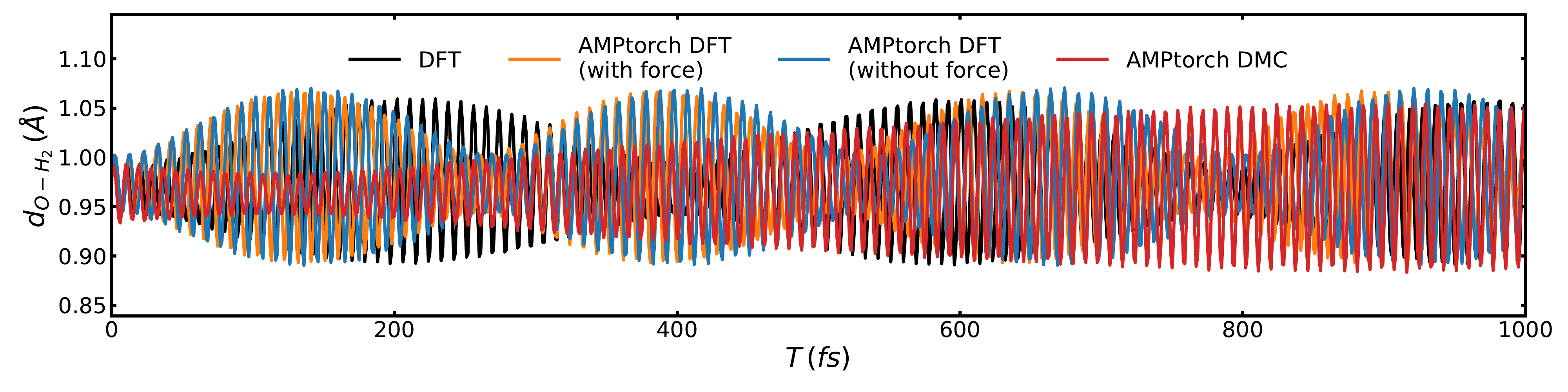}
    \label{fig:h2o_nve_oh2}
\end{subfigure}
\hfill
\begin{subfigure}[b]{\textwidth}
    \centering
    \includegraphics[width=\textwidth]{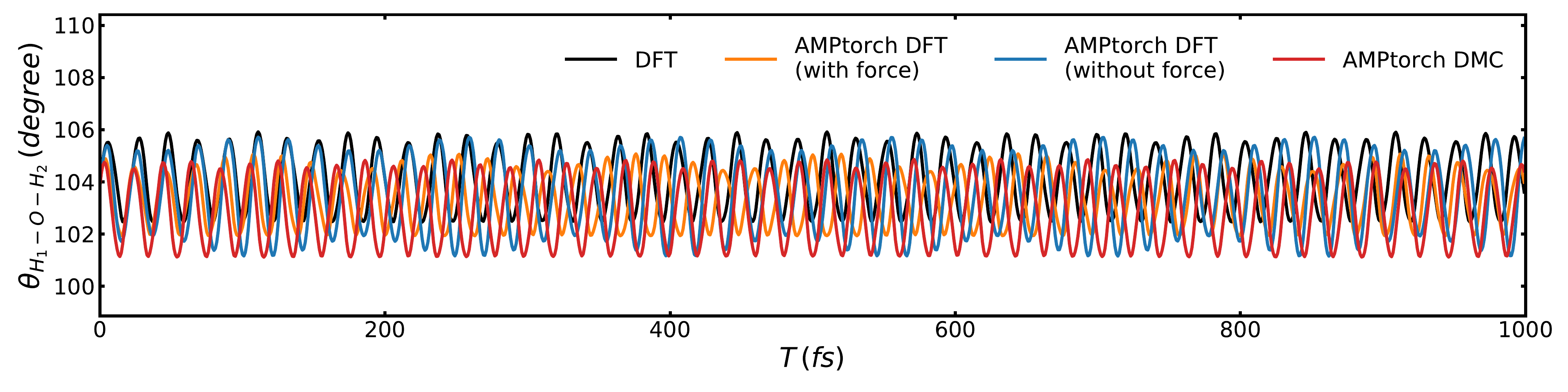}
    \label{fig:h2o_nve_theta}
\end{subfigure}
\caption{
    NVE molecular dynamics simulations for \ch{H2O} performed using DFT (black line), the AMPtorch DFT model with forces (orange line), the AMPtorch DFT model without forces (blue line), and the AMPtorch DMC model (red line). \textbf{(Top)} and \textbf{(Middle)}: O-H bond distances vs. time and \textbf{(Bottom)}: H-O-H bond angle vs. time. Here, we only plot 1000 fs of dynamics for illustration, but the full simulations extend beyond 2 ps, underscoring the stability of the trained AMPtorch-DFT and AMPtorch-DMC models.
}
\label{fig:h2o_nve}
\end{figure*}

From Figure \ref{fig:h2o_nve}, which depicts the predicted NVE dynamics of the water bond lengths and angle, \CC{we observe that the predicted amplitudes of the oscillations in the O-H bond distances and the H-O-H bond angle for the DFT and AMPtorch model dynamics are quite similar, with fluctuations remaining within a few percent of the average bond distance/angle. Despite the similar oscillation amplitudes, the longer-time dynamical envelopes (the groups of oscillations that grow and decay with their own unique frequencies) that can be seen to emerge on these plots differ among the four different models due to the small, yet intrinsic differences among the four different multidimensional PESs. For the DFT-related results, the long-time dynamical envelopes produced by the AMPtorch DFT models are shorter than those produced by the DFT model. Moreover, these envelopes are much longer for the AMPtorch DMC model than the DFT-related models. These discrepancies likely arise because of the subtle differences between the PESs (see Figure \ref{fig:h2o_validation_R} and \ref{fig:h2o_validation_theta}) that result in different redistributions of the energy among the multiple degrees of freedom present in the molecule that occur in the NVE ensemble. That is, even if only slight differences exist in the potential energy along each of the dimensions, these small errors can lead to much larger differences in the way the energy is distributed among degrees of freedom during a molecular dynamics simulation. Within the first 100 fs in Figure \ref{fig:h2o_nve} (e.g., $d_{O-H_{1}}$), the DFT and AMPtorch DFT models show almost identical trajectories, but over time, more and more subtle energy differences accumulate as a result of the energy redistribution, leading the $d_{O-H_{2}}$ trajectory for the AMPtorch DFT models to attain their largest amplitudes earlier than the DFT model, thus also resulting in a shorter long-time dynamical envelope. It should be noted that this is not affected by whether the AMPtorch models are trained with or without forces because both of them show different long-time envelope compared to the DFT model.}

\begin{table}[]
\centering
\begin{tabular}{ccccc}
    \hline
    Observables& DFT & \begin{tabular}[c]{@{}c@{}}AMPtorch DFT\\ (with forces)\end{tabular} & \begin{tabular}[c]{@{}c@{}}AMPtorch DFT\\ (without forces)\end{tabular} &
    \begin{tabular}[c]{@{}c@{}}AMPtorch\\ DMC\end{tabular} \\
    \hline
    \multicolumn{1}{c}{} & \multicolumn{4}{c}{NVE dynamics} \\
    \hline 
    $d_{O-H_1}$ (\r{A}) & 0.9743 & \CC{0.9788} & \CC{0.9791} & \CC{0.9685} \\
    $d_{O-H_2}$ (\r{A}) & 0.9739 & \CC{0.9789}  & \CC{0.9792} & \CC{0.9669} \\
    $d_{H_1-O-H_2}$ ($^\circ$) & 104.10 & \CC{103.35} & \CC{103.50} & \CC{102.92} \\
    \hline
    \multicolumn{1}{c}{} & \multicolumn{4}{c}{NVT dynamics} \\
    \hline
    $d_{O-H_1}$ (\r{A}) & 0.9695 & \CC{0.9743} & \CC{0.9748} & \CC{0.9643} \\
    $d_{O-H_2}$ (\r{A}) & 0.9708 & \CC{0.9758} & \CC{0.9764} & \CC{0.9656} \\
    $d_{H_1-O-H_2}$ ($^\circ$) & 104.44 & \CC{103.61} & \CC{103.98} & \CC{103.15} \\
    \hline
\end{tabular}
\caption{Mean \ch{H2O} bond distances and angles from the NVE and NVT simulations using DFT, the AMPtorch-DFT (with forces) model, the AMPtorch-DFT (without forces) model, and the AMPtorch-DMC model.}
\label{tab:h2o_nve_nvt}
\end{table}

This all noted, the AMPtorch DFT and AMPtorch DMC dynamics are stable and lead to meaningful average bond lengths and angles (see Table \ref{tab:h2o_nve_nvt}). First of all, even though the two O-H bonds were modeled independently, their MD averages end up being the same regardless of the model used, which serves as a check on our learning and simulations. In this case, all of the DFT models predict average bond lengths within \CC{0.005 \r{A} (0.5\%) and average bond angles within 0.8$^\circ$ (0.7\%)} of one another, which is excellent agreement. \CC{The AMPtorch DMC model predicts a shorter average bond length and a smaller average bond angle because the multidimensional minimum predicted by the AMPtorch DMC model is different than that predicted by the AMPtorch DFT models.} This results in oscillations that differ between the AMPtorch DMC and DFT models, which can also be observed in the geometry optimization results presented below (see Table \ref{tab:h2o_opt}). Because of the energy redistribution that ensues as described above, the average bond angles reported differ more dramatically from one another at the 2.5\% level, which is still respectable. The NVT average bond angles and bond lengths show similar trends (see Table \ref{tab:h2o_nve_nvt} and Supplementary Materials, but because the DMC bond length well is centered at shorter bond lengths, the DMC bond length predictions tend to be smaller. All of the models predict similar bond angles for NVT dynamics. These MD studies thus highlight the subtleties involved with ensuring correct, multidimensional dynamics, even when models seem to accurately predict potentials along individual dimensions. 

\paragraph*{Geometry Optimization}

\begin{figure}[t]
    \centerline{\resizebox{8cm}{!}{\includegraphics{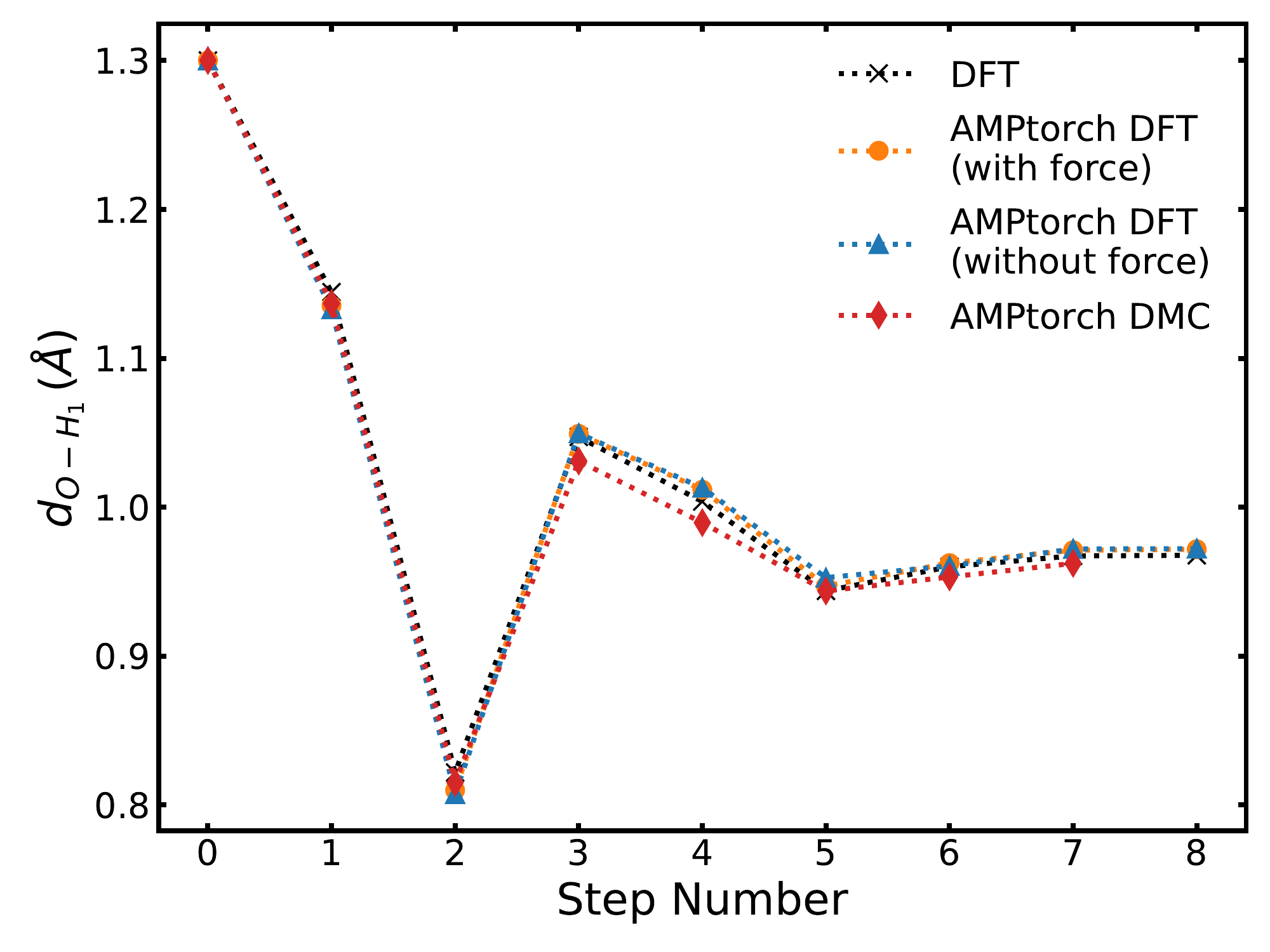}}}
    \caption{\ch{H2O} geometry relaxation using DFT, AMPtorch-DFT (with forces), AMPtorch-DFT (without forces), and AMPtorch-DMC. The optimizations using these models required \CC{8, 8, 8, and 7} iterations, respectively.}
    \label{fig:h2o_optimization}
  \end{figure}

As for \ch{C2}, geometry optimization was performed for \ch{H2O} using DFT and the AMPtorch models. For these optimizations, the initial \ch{H2O} structure was generated by stretching one O-H bond to 1.3 \r{A} and setting the bond angle to 120$^\circ$, while maintaining $d_{O-H_2}$ at its equilibrium length of 0.969 \r{A}. This initial structure serves as a compromise between being distorted and remaining within the training region. From Figure \ref{fig:h2o_optimization}, which shows how one of the O-H bond lengths changes during the optimization (see \CC{Tables S6 - S7} for more information about the convergence of $d_{O-H_2}$ and $\theta_{H_1-O-H_2}$), it is clear that this three-dimensional relaxation is more challenging, requiring more iterations than the optimization of \ch{C2}. Indeed, even fully \textit{ab initio} DFT required eight iterations to converge and did not converge in a smooth manner. \CC{However, we still observed that all three AMPtorch models are capable of optimizing the \ch{H2O} molecule following similar optimization paths as the DFT model. This said, the models trained with or without forces are able to converge their bond lengths and angles with less than a 1\% error within a reasonable number of iterations.}

\begin{table}[]
\centering
\begin{tabular}{ccccc}
    \hline
    \begin{tabular}[c]{@{}c@{}}Optimized\\ Observables\end{tabular}& DFT & \begin{tabular}[c]{@{}c@{}}AMPtorch DFT\\ (with forces)\end{tabular} & \begin{tabular}[c]{@{}c@{}}AMPtorch DFT\\ (without forces)\end{tabular} &
    \begin{tabular}[c]{@{}c@{}}AMPtorch\\ DMC\end{tabular} \\
    \hline
    $d_{O-H_1}$ & 0.9677 & \CC{0.9718} & \CC{0.9722} & \CC{0.9623} \\
    $d_{O-H_2}$ & 0.9674 & \CC{0.9719}  & \CC{0.9726} & \CC{0.9631} \\
    $d_{H_1-O-H_2}$ & 104.31 & \CC{103.52} & \CC{103.56} & \CC{102.83}  \\
    \hline
\end{tabular}
\caption{\ch{H2O} bond lengths and angles obtained after geometry relaxation using DFT, AMPtorch-DFT (with forces), AMPtorch-DFT (without forces), and AMPtorch-DMC. The optimizations using these models required \CC{8, 8, 8, and 7} iterations, respectively.}
\label{tab:h2o_opt}
\end{table}

\subsubsection{Case 3: \ch{CH3Cl} Molecule}\label{sec:ch3cl}

As a final illustration of our machine learning approach, we performed DMC-quality molecular dynamics simulations and geometry optimizations on methyl chloride, \ch{CH3Cl}, our most difficult example yet. With 9 degrees of freedom, \ch{CH3Cl} possesses a much higher dimensional PES than \ch{C2} and \ch{H2O}, which makes constructing a useful training set a substantially more difficult task than for the previous cases. Indeed, because of the high dimensional nature of the system, it is neither practical nor desirable to use grid sampling along each of the dimensions. For instance, even taking 10 grid points in each dimension would mean that $10^{9}$ \textit{ab initio} points would be needed for \ch{CH3Cl}, which would be virtually impossible to assemble.

Instead of grid sampling the PES along \ch{CH3Cl}'s degrees of freedom, the procedure to sample PES points presented by Owens \textit{et al.}\cite{owens2015accurate} was adopted to prepare the training dataset for \ch{CH3Cl}. In summary, Owens \textit{et al.} prepared their data set using a random energy-weighted Monte Carlo sampling algorithm for the nine internal coordinates, which leads to a global grid of 44,820 geometries with energies up to roughly 6 eV. Around 1,000 low-energy points were specifically added into the dataset to ensure the adequate description of the equilibrium region. The AMPtorch models were constructed using the same procedure as was used for \ch{C2} and \ch{H2O}, and subsequently employed for the molecular dynamics and optimization calculations. \CC{Details regarding the model construction and training results are included in the Supplementary Materials.}

\begin{table}[]
  \centering
  \begin{tabular}{ccccc}
      \hline
      Observables & DFT & \begin{tabular}[c]{@{}c@{}}AMPtorch DFT\\ (with forces)\end{tabular} & \begin{tabular}[c]{@{}c@{}}AMPtorch DFT\\ (without forces)\end{tabular} & \begin{tabular}[c]{@{}c@{}}AMPtorch\\DMC\end{tabular}  \\
      \hline
      \multicolumn{1}{c}{} & \multicolumn{4}{c}{NVE dynamics} \\
      \hline 
      $d_{C-Cl}$ (\r{A}) & 1.8111 & \CC{1.8113} & \CC{1.8125} & \CC{1.7957}\\
      $d_{C-H_1}$ (\r{A}) & 1.0986 & \CC{1.0983} & \CC{1.0988} & \CC{1.0875}\\
      $d_{C-H_2}$ (\r{A}) & 1.0957 & \CC{1.0956} & \CC{1.0960} & \CC{1.0867}\\
      $d_{C-H_3}$ (\r{A}) & 1.0962 & \CC{1.0965} & \CC{1.0989} & \CC{1.0868}\\
      $\theta_{Cl-C-H_1}$ ($^\circ$) & 108.44 & \CC{108.42} & \CC{108.40} & \CC{108.39}\\
      $\theta_{Cl-C-H_2}$ ($^\circ$) & 108.15 & \CC{108.12} & \CC{108.05} & \CC{108.07}\\
      $\theta_{Cl-C-H_3}$ ($^\circ$) & 108.24 & \CC{108.21} & \CC{108.46} & \CC{108.14}\\
      $\theta_{H_1-C-H_2}$ ($^\circ$) & 110.45 & \CC{110.47} & \CC{110.34} & \CC{110.50}\\
      $\theta_{H_1-C-H_3}$ ($^\circ$) & 110.46 & \CC{110.48} & \CC{110.54} & \CC{110.48}\\
      $\theta_{H_2-C-H_3}$ ($^\circ$) & 110.54 & \CC{110.58} & \CC{110.48} & \CC{110.67}\\
      \hline
      \multicolumn{1}{c}{} & \multicolumn{4}{c}{NVT dynamics} \\
      \hline 
      $d_{C-Cl}$ (\r{A}) & 1.7963 & \CC{1.7965} & \CC{1.7940} & \CC{1.7809} \\
      $d_{C-H_1}$ (\r{A}) & 1.0966 & \CC{1.0968} & \CC{1.0976} & \CC{1.0869}\\
      $d_{C-H_2}$ (\r{A}) & 1.0972 & \CC{1.0973} & \CC{1.0981} & \CC{1.0875}\\
      $d_{C-H_3}$ (\r{A}) & 1.0989 & \CC{1.0990} & \CC{1.0998} & \CC{1.0891}\\
      $\theta_{Cl-C-H_1}$ ($^\circ$) & 108.33 & \CC{108.30} & \CC{108.41} & \CC{108.22}\\
      $\theta_{Cl-C-H_2}$ ($^\circ$) & 108.32 & \CC{108.29} & \CC{108.40} & \CC{108.22}\\
      $\theta_{Cl-C-H_3}$ ($^\circ$) & 108.62 & \CC{108.59} & \CC{108.70} & \CC{108.48}\\
      $\theta_{H_1-C-H_2}$ ($^\circ$) & 110.46 & \CC{110.48} & \CC{110.38} & \CC{110.57}\\
      $\theta_{H_1-C-H_3}$ ($^\circ$) & 110.48 & \CC{110.50} & \CC{110.39} & \CC{110.59}\\
      $\theta_{H_2-C-H_3}$ ($^\circ$) & 110.10 & \CC{110.13} & \CC{110.01} & \CC{110.26}\\
      \hline
  \end{tabular}
  \caption{Mean \ch{CH3Cl} bond lengths and angles averaged over DFT, AMPtorch DFT (with forces), AMPtorch DFT (without forces), and AMPtorch DMC trajectories in the NVE and NVT ensembles.}
  \label{tab:ch3cl_nve_nvt}
\end{table}

\begin{table}[]
    \centering
    \begin{tabular}{ccccc}
        \hline        
        \begin{tabular}[c]{@{}c@{}}Optimized\\ Observables\end{tabular}
        & DFT &  \begin{tabular}[c]{@{}c@{}}AMPtorch DFT\\ (with forces)\end{tabular} & \begin{tabular}[c]{@{}c@{}}AMPtorch DFT\\ (without forces)\end{tabular} & \begin{tabular}[c]{@{}c@{}}AMPtorch\\ DMC\end{tabular} \\
        \hline
        $d_{C-Cl}$ (\r{A}) & 1.7977 & \CC{1.7979} & \CC{1.7919} & \CC{1.7807} \\
        $d_{C-H_1}$ (\r{A}) & 1.0939 & \CC{1.0940} & \CC{1.0953}  & \CC{1.0846} \\
        $d_{C-H_2}$ (\r{A}) & 1.0938 & \CC{1.0939} & \CC{1.0947} & \CC{1.0843} \\
        $d_{C-H_3}$ (\r{A}) & 1.0934 & \CC{1.0934} & \CC{1.0950} & \CC{1.0838} \\
        $\theta_{Cl-C-H_1}$ ($^\circ$) & 108.36 & \CC{108.32} & \CC{108.42} & \CC{108.32} \\
        $\theta_{Cl-C-H_2}$ ($^\circ$) & 108.32 & \CC{108.29} & \CC{108.44} & \CC{108.29} \\
        $\theta_{Cl-C-H_3}$ ($^\circ$) & 108.39 & \CC{108.37} & \CC{108.59} & \CC{108.33} \\
        $\theta_{H_1-C-H_2}$ ($^\circ$) & 110.66 & \CC{110.69} & \CC{110.60} & \CC{110.64} \\
        $\theta_{H_1-C-H_3}$ ($^\circ$) & 110.61 & \CC{110.64} & \CC{110.41} & \CC{110.64} \\
        $\theta_{H_2-C-H_3}$ ($^\circ$) & 110.42 & \CC{110.44} & \CC{110.30} & \CC{110.54} \\
        \hline
    \end{tabular}
    \caption{Optimized bond distances and angles for \ch{CH3Cl} using DFT, the AMPtorch-DFT (with forces) model, the AMPtorch-DFT (without forces) model, and the AMPtorch-DMC model. These models required \CC{10, 10, 9, and 10} optimization iterations, respectively.}
    \label{tab:ch3cl_opt}
  \end{table}

\paragraph*{Molecular Dynamics Simulations}

\begin{figure*}[]
  \begin{subfigure}{\textwidth}
      \centering
      \includegraphics[width=\linewidth]{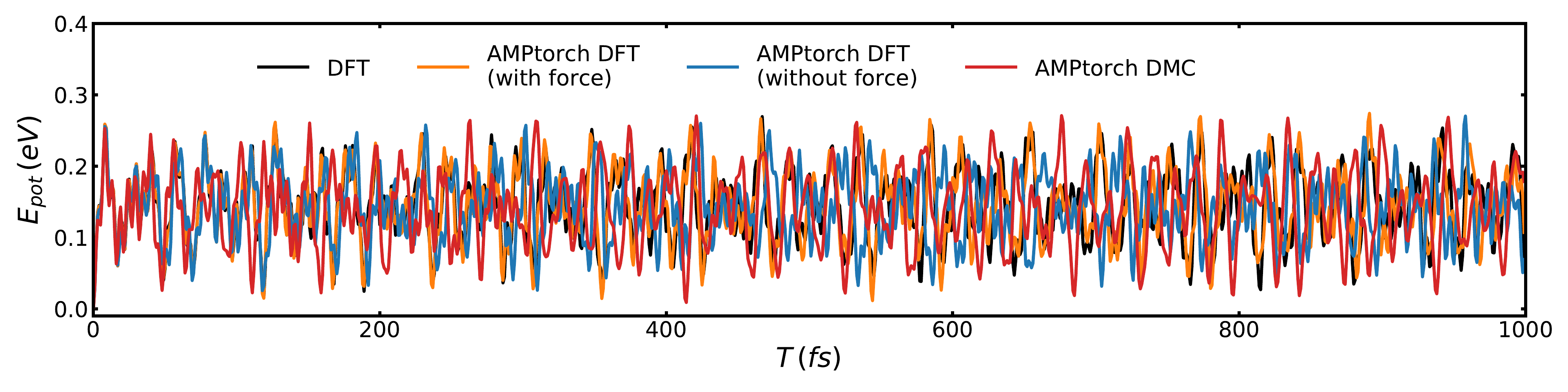}
  \end{subfigure}
  \begin{subfigure}{\textwidth}
      \centering
      \includegraphics[width=\linewidth]{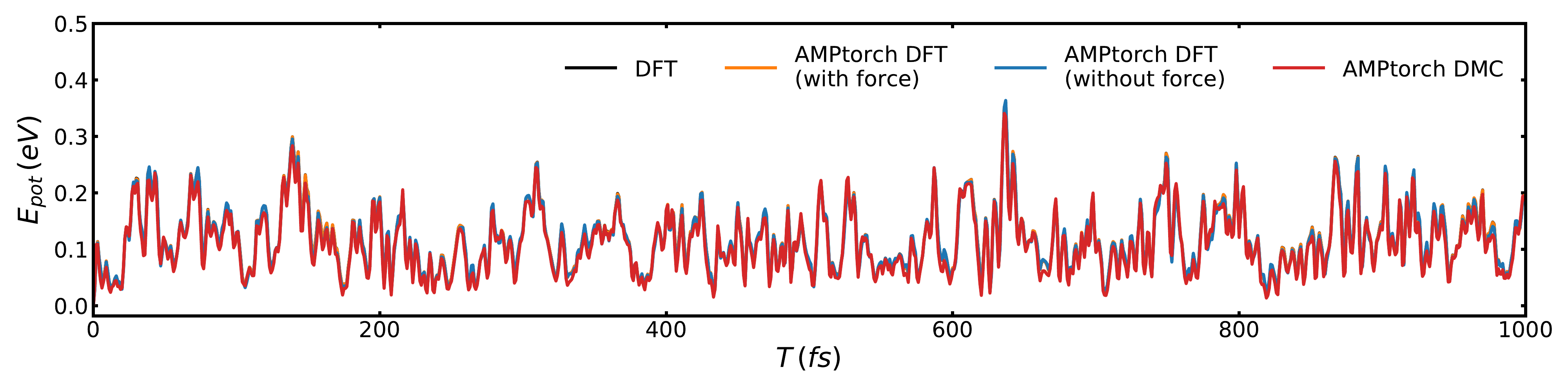}
  \end{subfigure}
  \caption{\ch{CH3Cl} potential energy vs. time plots for MD simulations performed using DFT (black line), AMPtorch DFT with forces (orange line), AMPtorch DFT without forces (blue line), and AMPtorch DMC (red line). \textbf{(Top)}: NVE E$_{\text{pot}}$ vs. time; and \textbf{(Bottom)}: NVT E$_{\text{pot}}$ vs. time. Here, we only present the first ps of the trajectories for illustration.}
   \label{fig:ch3cl_energies}
\end{figure*}

Because it is difficult to adequately represent the dynamics of all nine of \ch{CH3Cl}'s degrees of freedom, here, we instead focus on analyzing how the molecule's potential energies and average bond lengths and angles behave. In Figure \ref{fig:ch3cl_energies}, we plot the potential energies obtained from all four models over time for simulations in both the NVE (top) and NVT (bottom) ensembles (plots of the total energies and total kinetic energies are shown in Supplementary Materials). From the top panel, we see that, for the most part, all models seem to conserve energy, exhibiting the same amplitude fluctuations over time. While energy is not conserved in the NVT ensemble (see Figure \ref{fig:ch3cl_energies} bottom), the NVT potential energy curves show the same features in that the AMPtorch models \CC{manifest fluctuations similar to those exhibited by the \textit{ab initio} DFT simulations. Remarkably, the models without forces exhibit the same trends for bond lengths and bond angles as the models with forces.}

In Tables \ref{tab:ch3cl_nve_nvt}, we present the average bond lengths and angles obtained over the course of the NVE and NVT simulations. \CC{The AMPtorch DFT with forces results for bond lengths and angles are within 0.04\% of the DFT results across the board, while the forceless averages deviate by around 0.1\%, which is still noteworthy as a relatively small error value for a high-dimensional surface.} It can also be observed from this table that the AMPtorch models were able to maintain the symmetry of the molecule and correctly predict the energies of its symmetric structures despite the fact that the training set possessed asymmetric structures. 

\paragraph*{Geometry Optimization}

Geometry optimization was also performed for \ch{CH3Cl} using DFT and the AMPtorch models previously described. All structures relaxed were initialized to their minimum potential energy structures. Then, each atom was displaced based on a normal distribution with a standard deviation of 0.1 \r{A}. This displacement was performed using the rattle function implemented in the ASE package. After the displacements, the BFGS optimizer was used for geometry optimization of the \ch{CH3Cl} molecule using the forces generated by the four models. The convergence criterion was set as $f_{max} < 0.05$. It took DFT 10 iterations, the AMPtorch DFT with forces model \CC{10} iterations, AMPtorch DFT without forces model \CC{9} iterations, and the AMPtorch DMC model \CC{10} iterations to reach the convergence. \CC{All four models yielded very similar geometries. The bond length and bond angle differences between the AMPtorch models and the DFT model are well within 0.005 \r{A} and $0.1^\circ$ regardless of whether forces are included in the training. Optimization details are provided in the Supplementary Materials.}

\section{Conclusion}
\label{sec:conclusion}
In summary, in this work, we have employed machine learning to learn DMC forces from DMC potential energy data and to leverage those forces to perform DMC-quality molecular dynamics and geometry optimization calculations. DMC forces were learned using Behler-Parinello Neural Network models (BPNN) within the Atomistic Modeling Package (AMP) and used to model the dynamics of three illustrative molecules: \ch{C2} (one degree of freedom), \ch{H2O} (three degrees of freedom), and \ch{CH3Cl} (nine degrees of freedom). Although not perfect, we find that our machine learning models can learn forces with sufficient accuracy to perform geometry optimizations and yield reasonable estimates of average bond lengths and angles in molecular dynamics simulations at a cost hundreds of times less than the cost of \textit{ab initio} molecular dynamics simulations based on DMC energies and forces. Reproducing the exact dynamics of molecules without explicit force information is nevertheless found to be much more challenging given the dynamics' sensitivity to the precise curvature of potential energy landscapes. This work thus illustrates the potential - and limitations - of training force fields for dynamics calculations without explicit knowledge of forces, knowledge that is valuable whenever one is interested in accelerating \textit{ab initio} dynamics calculations based upon electronic structure methods that either cannot readily generate forces or can only do so at a steep cost. 

Although the learning techniques we employed here were sufficient for modeling small molecules, the fact that our training sets must grow combinatorially with the number of degrees of freedom to fully sample the potential energy surface ultimately limits our ability to generalize our approach to substantially larger species with many more degrees of freedom. One particularly promising way of avoiding the steep scaling that accompanies such grid-based sampling techniques is to instead employ active learning techniques, in which one begins training with a comparatively small training set and then iteratively adds training points to the set based upon uncertainties in specific regions of the potential energy landscape.\cite{shuaibi2020enabling} Indeed, Troy \textit{et al.}\cite{loeffler2020active} were able to successfully train a neural network to model water clusters using active learning with just 426 structures. Smith \textit{et al.}\cite{smith2018less} also employed active learning via Query by Committee, demonstrating that only 10 \% to 25\% of their COMP6 benchmark data set of organic molecules was needed to accurately represent the chemical space of the molecules. Our own initial explorations into active learning forces without initial force data have shown similar improvement, demonstrating the feasibility of extending similar approaches to substantially larger species, which will be the subject of a subsequent manuscript. 

Active learning techniques also open the door to a more compromising and likely more accurate approach to learning DMC forces. While this work focused on learning dynamics without \textit{any} knowledge of DMC forces, such learning can be viewed as an extreme case. Instead, one can take a more moderate stance of overwhelmingly training on less costly energy data and supplementing this training with select, but more expensive force data, as needed and available. This can be accomplished by training on a robust set of energy data and then using active learning to selectively incorporate not only energy, but force, information to improve predictions where that improvement is most needed. Such an approach could substantially decrease the amount of energy data needed at the cost of relatively few force calculations. Just how many force calculations would be needed to achieve set accuracy - and hence what the optimal ratio of energy to force calculations is for machine learning methods - remains an interesting question for further research. Indeed, even if QMC techniques that can more expediently compute forces for a wide class of molecules and materials emerge, the questions explored in this manuscript and the question of how limited force data can best be used to learn would still remain important for further accelerating \textit{ab initio} molecular dynamics simulations.

\begin{acknowledgement}

C.H. thanks Cheng Zeng and Muhammed Shuaibi for help with the AMPtorch calculations, and Gopal Iyer and Yichen Chai for insightful discussions regarding machine learning algorithms. B.R. graciously acknowledges financial support for this project from the Research Corporation of America, U.S. National Science Foundation grant OIA-1921199, and AFOSR Award Number FA9550-19-1-9999. C.H.'s work was supported by U.S. Department of Energy, Office of Science, Basic Energy Sciences Award \#DE-FOA-0001912. This research was conducted using computational resources and services at the Center for Computation and Visualization, Brown University.

\end{acknowledgement}

\begin{suppinfo}

The following files are available free of charge.
\begin{itemize}
  \item supplemental.pdf: A PDF document containing tables and plots of the MAEs, results for NVE and NVT dynamics, and geometry optimization data.
  \item The Python code used for our AMPtorch models, as well as performing MD simulation and structure optimization are available on \href{https://github.com/cancan233/ml_qmc_force}{https://github.com/cancan233/ml\_qmc\_force}.\cite{mldmcforces}
\end{itemize}

\end{suppinfo}

\bibliography{ml_dmc_force}

\end{document}